\documentclass[11pt]{article}%
\usepackage{natbib}
\usepackage{amsfonts}
\usepackage{amssymb}
\usepackage{amsmath}
\usepackage{lscape}
\usepackage{color}
\usepackage[left=1.7cm, right=2.5cm, top=1.7cm, bottom=1.7cm]{geometry}
\usepackage{graphicx}
\usepackage{booktabs}
\usepackage{xcolor}
\usepackage{algorithm}
\usepackage{algorithmic}
\usepackage{bm}%
\providecommand{\U}[1]{\protect\rule{.1in}{.1in}}
\definecolor{DarkO}{cmyk}{0.0, 0.43, 0.90, 0.13}

\def\bt{\boldsymbol{\theta }}
\DeclareMathOperator*{\argmax}{arg\,max}
\DeclareMathOperator*{\argmin}{arg\,min}
\begin{document}

\title{Bayesian Forecasting in Economics and Finance: A Modern Review}
\author{Gael M. Martin, David T. Frazier, Worapree Maneesoonthorn, Rub\'{e}n
Loaiza-Maya, \medskip
\and Florian Huber, Gary Koop, John Maheu, Didier Nibbering and Anastasios
Panagiotelis\thanks{{\footnotesize Authors: Gael M. Martin, Monash University,
Australia} ({\footnotesize Corresponding author: gael.martin@monash.edu);
David T. Frazier, Monash University, Australia; Worapreee (Ole)
Maneesoonthorn, Monash University, Australia; Rub\'{e}n Loaiza-Maya, Monash
University, Australia; Florian Huber; University of Salzburg, Austria; Gary
Koop, University of Strathclyde, UK; John Maheu, McMaster University, Canada;
Anastasios Panagiotelis; University of Sydney, Australia; Didier Nibbering,
Monash University, Australia. The authors would like to thank the Editor, an
associate editor and two anonymous referees for very helpful and constructive
comments on an earlier draft of the paper. Martin, Frazier and Maneesoonthorn
have been supported by Australian Research Council Discovery Grant
DP200101414. Frazier has also been supported by Australian Research Council
Discovery Early Career Researcher Award DE200101070}, {\footnotesize and
Loaiza-Maya supported by Australian Research Council Discovery Early Career
Researcher Award DE230100029. Huber has been supported by Austrian Science
Fund Grant ZK-35. Maheu is grateful for financial support from the Social
Sciences and Humanities Research Council of Canada.}} }
\maketitle

\begin{abstract}
The Bayesian statistical paradigm provides a principled and coherent approach
to probabilistic forecasting. Uncertainty about all\textit{ }unknowns that
characterize any forecasting problem -- model, parameters, latent states -- is
able to be quantified explicitly, and factored into the forecast distribution
via the process of integration or averaging. Allied with the elegance of the
method, Bayesian forecasting is now underpinned by the burgeoning field of
Bayesian computation, which enables Bayesian forecasts to be produced for
virtually any problem, no matter how large, or complex. The current state of
play in Bayesian forecasting in economics and finance is the subject of this
review. The aim is to provide the reader with an overview of modern approaches
to the field, set in some historical context; and with sufficient
computational detail given to assist the reader with implementation.

\bigskip


\bigskip

\end{abstract}

\newpage

\section{Introduction}

\subsection{Why Bayesian forecasting?}

\baselineskip17.5pt

The Bayesian statistical paradigm uses {the rules and language of probability}
to quantify uncertainty about all unknown aspects of phenomena that generate
observed data. This core characteristic of the paradigm makes it particularly
suitable for forecasting, with uncertainty about the unknown values of future
observations automatically expressed in terms of a probability distribution.
Moreover, Bayesian methods -- in principle\textbf{ -- }allow a user to
seamlessly, and systematically, yield probabilistic forecasts that reflect
uncertainty about all unknowns and that, as a consequence, condition primarily
on \textit{known} past events,\textbf{ }or \textit{data}: a feature that
\cite{geweke2006bayesian} refer to as the \textit{principle of relevant
conditioning}.

Indeed, the ability of Bayesian forecasters to appropriately incorporate the
uncertainty associated with the production of forecasts, while utilizing all
available information -- both \textit{a priori} and sample information\textbf{
--} in a principled manner, led \cite{granger1986forecasting} to conclude
that: \medskip

\begin{center}
\textit{\textquotedblleft In terms of forecasting accuracy a good Bayesian
will beat a non-Bayesian, }\newline\textit{who will do better than a bad
Bayesian.\textquotedblright} \medskip
\end{center}

\noindent Echoing these sentiments, in our opinion, the power of the Bayesian
forecasting paradigm is a product of the paradigm's ability to treat all
elements of the statistical problem necessary to produce forecasts -- future
observations, past observations, parameters, latent variables, models
--\textbf{ }as arguments\textbf{ }of a joint probability distribution. The
express probabilistic formulation of these elements, in turn, allows a
Bayesian to invoke the standard rules of probability to produce a distribution
for an unknown future value that is conditioned on the known past data, and is
\textit{marginal} of other arguments that are inherently unknown.

While this ability to marginalize all unknowns through probability calculus is
the hallmark of the Bayesian approach, the benefits of the paradigm, and what
ultimately in our opinion defines a `good Bayesian', is the attention to
detail necessary to successfully implement Bayesian methods. In Bayesian
forecasting, before we ever attempt to produce a forecast, we must first
carefully enumerate all possible sources of uncertainty -- including, where
possible, the set of alternative forecasting models; and construct reasonable
prior beliefs for these quantities, which often include (possibly several
layers of) latent variables that have a specific and delicate interaction with
the observed data; always taking great care to ensure that these prior beliefs
do not conflict with the observed data. Then and only then can we `turn the
Bayesian crank'\ to produce the joint posterior distribution over all unknown
quantities (including future values), and ultimately\textbf{ }integrate out
the quantities we are not interested in to obtain the (posterior) predictive
distribution for the future\textbf{ }values of our random variables of
interest. The attention to detail necessary to produce Bayesian forecasts aims
to reduce the number of \textit{implicit} maintained\textbf{ }assumptions, and
what explicit assumptions are maintained (e.g. the conditioning on a
particular model, or finite model set) can often be rationalized/tested
against the data.

Consistent with the internal coherence of the Bayesian statistical paradigm,
the basic manner in which all Bayesian forecasting problems are
\textit{framed} is the same. What differs however, from case to case, is the
way in which the problem is \textit{solved} -- i.e. the way in which the
forecast distribution is \textit{accessed. }To understand why this is so, it
is sufficient to recognize that virtually all Bayesian quantities of interest,
including forecast distributions, can be expressed as expectations of some
sort. For most models that are used to predict empirically relevant data these
expectations are not available in closed form. Hence, in any practical
problem, implementation of Bayesian forecasting is both model- and
data-dependent, and relies on advanced computational tools. Different
forecasting problems -- defined by different forms and `sizes' of models and
data sets -- require, in turn, different approaches to computation. The
evolution of the practice of Bayesian forecasting has, as a consequence, gone
hand-in-hand with developments on the computational front; with increasingly
large and complex models rendered amenable to a Bayesian forecasting approach
via access to modern techniques of computation.

\subsection{The purview of this review}

In this review, we give a modern take on the current landscape of Bayesian
forecasting. Whilst excellent textbook treatments of Bayesian forecasting are
given in \cite{geweke2005contemporary} and \cite{west2006bayesian}, and with
\cite{geweke2006bayesian} reviewing specific aspects of Bayesian forecasting
in a slew of practical settings, the field has advanced by leaps and bounds in
the last twenty years. Therefore, we believe the time is ripe to consider a
review of the subject that touches on many of the novel and exciting areas now
being explored. The methodological advances we review have general
applicability to all discipline areas; nevertheless, due to our own interests,
expertise and experience -- and to keep the scope of the paper manageable --
we have chosen to focus primarily on applications in the economic sciences.
Whilst the paper is not designed to be a treatise on Bayesian computation,
sufficient details are provided to enable the practitioner to understand
\textit{why}\textbf{ }numerical tools are needed in most forecasting settings,
and \textit{how }they are used.

The general structure of the paper is as follows. In Section \ref{primer} we
provide a short tutorial on Bayesian forecasting. This begins with an outline
of the Bayesian forecasting method, followed by an overview of the
computational techniques used to \textit{implement }the method. In Section
\ref{hist_persp} we then take the reader on a potted chronological tour of
Bayesian forecasting, up to the present day. We begin by giving a snapshot of
the forecasting problems tackled during the last decade of the 20th century
(and the early years of the 21st), and the computational solutions that were
adopted then -- most notably, Markov chain Monte Carlo (MCMC) algorithms. We
then look at the types of `intractable' forecasting problems that are
increasingly encountered in the 21st century, and provide an overview of the
new computational solutions that have been proposed to tackle such problems.
We also outline very recent developments in which misspecification of the
forecasting model is explicitly acknowledged, and conventional
likelihood-based Bayesian forecasting eschewed as a consequence; with
problem-specific measures of forecast accuracy (or forecast loss) used,
instead, to drive the production of forecast distributions. Section
\ref{reviews} then provides the reader with more detailed reviews of
contemporary Bayesian forecasting in the following four broad fields:
macroeconomics, finance, marketing, and electricity pricing and demand.
Section \ref{concl} closes the paper with a brief summary of the current state
of play.

Before proceeding further, we make a note about scope and language. To render
the scope of the paper manageable we focus primarily on Bayesian forecasting
in `time series models' -- i.e. models for random variables that are indexed
by time -- and on using such models to say something about the values that
these random variables will assume in the future. These future values may be
informed only by past observations\ on the variable, or may also depend on the
known values of covariates, or regressors. We also follow the convention in
the Bayesian literature by using the terms `forecast' and `prediction' (and
all of their various grammatical derivations) synonymously and interchangeably
in this case, for the sake of linguistic variety. The fundamental principles
of Bayesian prediction apply equally to data indexed by something other than
time. The term `forecast' is not used in this case as it is a term reserved
for temporal settings. The main exceptions to our focus on time series models,
and forecasting \textit{per se}, occur in Section \ref{mark}, in which models
for cross-sectional data are used to predict customer choice in marketing
settings, and Section \ref{elect}, in which models for electricity demand that
have a spatial dimension are referenced.

\section{A Tutorial on Bayesian Forecasting\label{primer}}

\subsection{The Bayesian forecasting method\label{tute}}

For the sake of illustration, we assume a scalar random variable $y_{t}$, and
define the $(T\times1)$ vector of observations on $y_{t}$ as $\mathbf{y}%
_{1:T}=(y_{1},y_{2},...,y_{T})^{\prime}$. We assume (for the moment) that
$\mathbf{y}_{1:T}$ has been generated from some parametric model with
likelihood $p(\mathbf{y}_{1:T}|\boldsymbol{\theta})$, with $\boldsymbol{\theta
}=(\theta_{1},\theta_{2},...,\theta_{p})^{\prime}$ ${\in\Theta\subseteq}$
$\mathbb{R}^{p}$ a $p$-dimensional vector of unknown parameters, and where we
possess prior beliefs on $\boldsymbol{\theta}$ specified by
$p(\boldsymbol{\theta})$. Using the same symbol $\mathbf{y}_{1:T}$ to denote
both the vector of observed data and the $T$-dimensional vector random
variable, we define the joint distribution over $\mathbf{y}_{1:T}$ and
$\boldsymbol{\theta}$ as $p(\mathbf{y}_{1:T},\boldsymbol{\theta}).$
Application of the standard rules of probability to $p(\mathbf{y}%
_{1:T},\boldsymbol{\theta})$ yields \textit{Bayes theorem }(or \textit{Bayes
rule)},%
\begin{equation}
p(\boldsymbol{\theta}\mathbf{|y}_{1:T})=\frac{p(\mathbf{y}_{1:T}%
|\boldsymbol{\theta})p(\boldsymbol{\theta})}{p(\mathbf{y}_{1:T})},
\label{posterior_pdf}%
\end{equation}
where $p(\mathbf{y}_{1:T})=\int_{{\Theta}}p(\mathbf{y}_{1:T}%
|\boldsymbol{\theta})p(\boldsymbol{\theta})d\boldsymbol{\theta}$. Bayes
theorem provides a representation for the \textit{posterior} probability
density function (pdf) for $\boldsymbol{\theta}$, $p(\boldsymbol{\theta
}\mathbf{|y}_{1:T})$, as proportional to the product of the likelihood
function and the prior. The term $p(\mathbf{y}_{1:T})$ defines the marginal
likelihood, and the scale factor $\left[  p(\mathbf{y}_{1:T})\right]  ^{-1}$
in (\ref{posterior_pdf}) ensures that $p(\boldsymbol{\theta}|\mathbf{y}%
_{1:T})$ integrates to one.

Now, define $y_{T+1}$ as the (one-step-ahead) future random variable, where we
focus on one-step-ahead forecasting in Sections \ref{primer} and
\ref{hist_persp} merely to simplify the exposition. Assuming $y_{T+1}$ to be a
continuous random variable (again, for illustration), standard probability
manipulations lead to the following expression for the \textit{forecast} (or
\textit{predictive}) pdf for $y_{T+1}:$%
\begin{equation}
p(y_{T+1}|\mathbf{y}_{1:T})=\int_{{\Theta}}p(y_{T+1}|\boldsymbol{\theta
}\mathbf{,y}_{1:T})p(\boldsymbol{\theta}\mathbf{|y}_{1:T})d\boldsymbol{\theta
}. \label{exact_pred}%
\end{equation}
When no confusion arises, we also refer to $p(y_{T+1}|\mathbf{y}_{1:T})$,
albeit loosely, as the forecast (or predictive) \textit{distribution}, or
simply as the `predictive'.\footnote{We note that $p(y_{T+1}|\mathbf{y}%
_{1:T})$ is sometimes referred to as a `posterior' predictive in the
literature, given that it is produced by averaging the conditional predictive,
$p(y_{T+1}|\boldsymbol{\theta}\mathbf{,y}_{1:T})$, with respect to the
posterior density, $p(\boldsymbol{\theta}\mathbf{|y}_{1:T}).$ We do not adopt
this expression, leaving it to the context to make it clear as to whether the
term `predictive' is being used to refer to the distribution that is marginal
of $\boldsymbol{\theta}$, $p(y_{T+1}|\mathbf{y}_{1:T})$, or that which is
conditioned on $\boldsymbol{\theta}$, $p(y_{T+1}|\boldsymbol{\theta
}\mathbf{,y}_{1:T})$. We also streamline the exposition by not using explicit
notation for any observed covariates on which the model for $y_{t}$ may
depend, and on which the predictive for $y_{T+1}$ would condition, unless this
is essential.} The density $p(y_{T+1}|\mathbf{y}_{1:T})$ summarizes all
uncertainty about $y_{T+1}$, conditional on the assumed model -- which
underpins the structure of both the conditional predictive, $p(y_{T+1}%
|\boldsymbol{\theta}\mathbf{,y}_{1:T})$, and the posterior itself -- and the
prior beliefs that inform $p(\boldsymbol{\theta}\mathbf{|y}_{1:T}).$ Point and
interval predictions of $y_{T+1}$, and indeed any other distributional
summary, can be extracted from (\ref{exact_pred}). In the case where the model
itself is uncertain, and a finite set of parametric models, $\mathcal{M}_{1}$,
$\mathcal{M}_{2}$,...,$\mathcal{M}_{K},$ is assumed to span the model space, a
`model-averaged' predictive (e.g. \citealp{raftery:madigan:hoeting:1997},
Section 2), $p_{MA}(y_{T+1}|\mathbf{y}_{1:T})$, is produced as
\begin{equation}
p_{MA}(y_{T+1}|\mathbf{y}_{1:T})=\sum\limits_{k=1}^{K}p(y_{T+1}|\mathbf{y}%
_{1:T},\mathcal{M}_{k})p(\mathcal{M}_{k}|\mathbf{y}_{1:T})\text{,}
\label{mod_av}%
\end{equation}
where $p(y_{T+1}|\mathbf{y}_{1:T},\mathcal{M}_{k})$ denotes the density in
(\ref{exact_pred}), but now conditioned explicitly on the $kth$ model in the
set. The $kth$ posterior model probability, $p(\mathcal{M}_{k}|\mathbf{y}%
_{1:T})$, $k=1,2,...,K,$ is computed via a further application of Bayes
theorem in which the (initial) joint distribution of interest is defined over
both the model space and the space for the parameters of each of the $K$
models. Standard manipulations lead to%
\begin{equation}
p(\mathcal{M}_{k}|\mathbf{y}_{1:T})\propto p(\mathbf{y}_{1:T}|\mathcal{M}%
_{k})p(\mathcal{M}_{k}), \label{post_mod}%
\end{equation}
where
\begin{equation}
p(\mathbf{y}_{1:T}|\mathcal{M}_{k})=\int_{{\Theta}_{k}}p(\mathbf{y}%
_{1:T}\mathbf{|}\boldsymbol{\theta}_{k},\mathcal{M}_{k})p(\boldsymbol{\theta
}_{k}|\mathcal{M}_{k})d\boldsymbol{\theta}_{k}, \label{marg_k}%
\end{equation}
for each $k=1,2,...,K$, with $\boldsymbol{\theta}_{k}$ denoting the parameter
set for the $kth$ model.

As is clear, \textit{analytical }evaluation of $p(y_{T+1}|\mathbf{y}_{1:T})$
in (\ref{exact_pred}) requires, at the very least, a closed-form expression
for $p(\boldsymbol{\theta}\mathbf{|y}_{1:T})$. Typically, however, such an
expression is not available, with most posteriors being known only up to a
constant of proportionality, as
\begin{equation}
p(\boldsymbol{\theta}\mathbf{|y}_{1:T})\propto p(\mathbf{y}_{1:T}%
|\boldsymbol{\theta})p(\boldsymbol{\theta}). \label{Bayes_proport}%
\end{equation}
The main exceptions to this occur when $p(\mathbf{y}_{1:T}|\boldsymbol{\theta
})$ is from the exponential family, and either a natural conjugate, or
convenient noninformative prior is adopted; specifications which may be
suitable for some simple (and low-dimensional) empirical problems, but are
certainly not broadly applicable in practice. Analytical evaluation of
$p_{MA}(y_{T+1}|\mathbf{y}_{1:T})$ in (\ref{mod_av}) also requires a
closed-form expression for each $p(\mathbf{y}_{1:T}|\mathcal{M}_{k})$ (with
normalization of $p(\mathcal{M}_{k}|\mathbf{y}_{1:T})$ then straightforward);
once again a rare thing beyond the exponential family (and standard prior)
setting. Hence the need for numerical \textit{computation }to implement
Bayesian forecasting in virtually all realistic empirical
problems.\footnote{Numerous textbook illustrations of the material in this
section can be found. In addition to \cite{geweke2005contemporary} and
\cite{west2006bayesian} as cited above, some examples are \cite{zellner:1971},
\cite{koop2003bayesian} and \cite{robert:2007}. We also refer the reader to
\cite{steele:2020} for a recent review of Bayesian model averaging in
economics.}

\subsection{An overview of computation\label{comp}}

The form of (\ref{exact_pred}) makes it clear that the Bayesian predictive
pdf, $p(y_{T+1}|\mathbf{y}_{1:T})$, is nothing more than the posterior
expectation of the predictive that conditions\textit{ }on $\boldsymbol{\theta
}$. Hence, accessing $p(y_{T+1}|\mathbf{y}_{1:T})$ amounts to the evaluation
of an expectation. This insight is helpful, as it enables us to see many of
the computational methods that are used to access $p(y_{T+1}|\mathbf{y}%
_{1:T})$ -- in cases where it is not available in closed form -- simply as
different ways of numerically estimating an expectation.

It is convenient to group Bayesian computational methods into three
categories: (i) deterministic integration (or quadrature) methods
(\citealp{davis1975numerical}; \citealp{naylor:smith:1982}); (ii) exact
simulation methods; and (iii) approximate methods. Given that the production
of $p(y_{T+1}|\mathbf{y}_{1:T})$ involves integration over $\boldsymbol{\theta
}$, only in very low-dimensional models is (i) a feasible computational
approach on its own, due to the well-known `curse of dimensionality' that
characterizes numerical quadrature. Hence, the computational methods in (ii)
and (iii) are those most commonly adopted, and will be our focus here; noting
that quadrature may play still a limited role \textit{within} these
alternative computational frameworks.

The methods in (ii)\textit{ }use simulation\textit{\ }to produce $M$ draws of
$\boldsymbol{\theta}$, $\boldsymbol{\theta}^{(i)}$, $i=1,2,...,M$, from the
posterior $p(\boldsymbol{\theta}\mathbf{|y}_{1:T})$, which, in turn, define
$M$ conditional predictives, $p(y_{T+1}|\boldsymbol{\theta}^{(i)}%
\mathbf{,y}_{1:T})$, $i=1,2,...,M$, the mean of which is used to estimate
(\ref{exact_pred}). Alternatively, if it is easier to simulate from
$p(y_{T+1}|\boldsymbol{\theta}^{(i)}\mathbf{,y}_{1:T})$ than to evaluate it at
any point in the support of $y_{T+1}$, $M$ draws of $y_{T+1}$, $y_{T+1}^{(i)}%
$, $i=1,2,...,M$, are taken, one for each draw $\boldsymbol{\theta}^{(i)}$,
and kernel density estimation methods used to produce an estimate of
$p(y_{T+1}|\mathbf{y}_{1:T})$. Different simulation methods are distinguished
by the way in which the posterior draws are produced. Methods in (ii)\textit{
}include Monte Carlo sampling (\citealp{metropolis:ulam:1949}), importance
sampling (IS) (\citealp{hammersley:handscomb:1964};
\citealp{kloek1978bayesian}; \citealp{geweke:1989}) and MCMC sampling --
including Gibbs sampling (\citealp{geman:1984}; \citealp{gelfand:smith90}) and
Metropolis-Hastings (MH) algorithms (\citealp{metropolis:1953};
\citealp{hastings:1970}) -- with MCMC being by far the most common simulation
method used to compute forecast distributions in practice. The term `exact'
arises from the fact that, under appropriate conditions (including convergence
of the Markov chain to $p(\boldsymbol{\theta}\mathbf{|y}_{1:T})$ in the case
of the MCMC algorithms), such methods all produce a $\sqrt{M}$-consistent
estimate of the ordinate $p(y_{T+1}|\mathbf{y}_{1:T})$, at any point in the
support of the random variable $y_{T+1}$; this estimate can thus be rendered
arbitrarily accurate, for large enough $M.$

We refer the reader to: \cite{chib_greenberg_1996} and
\cite{geyer2011introduction} for reviews of MCMC sampling;
\cite{casella:george:1992} and \cite{chibandgreenberg:1995} for descriptions
of the Gibbs and MH algorithms (respectively) that are useful for
practitioners; and \cite{andrieu2004computational}, \cite{robert:casella:2011}
and \cite{martin2022history}\textbf{ }for historical accounts of MCMC
sampling. \cite{geweke2006bayesian} also serves as an excellent reference on
the use of these computational methods in a forecasting context. Given the
critical role played by MCMC methods in the production of Bayesian forecasts,
the basic principles of the algorithms are also outlined below in Section
\ref{sol}; with more recent developments of both IS and MCMC -- most notably
sequential Monte Carlo (SMC) (\citealp{gordon:salmon:smith:1993};
\citealp{Chopin2020}) and pseudo-marginal MCMC (\citealp{beaumont:2003};
\citealp{andrieu:roberts:2009}; \citealp{andrieu:doucet:holenstein:2010})\ --
discussed briefly in Section \ref{intrac}.

The methods in (iii)\textit{ }replace $p(\boldsymbol{\theta}\mathbf{|y}%
_{1:T})$ in the integrand of (\ref{exact_pred}) with an \textit{approximation}
of some sort, and evaluate the resultant integral. In so doing, such methods
do not aim to estimate $p(y_{T+1}|\mathbf{y}_{1:T})$ itself, but some
representation of it, defined as the expectation of $p(y_{T+1}%
|\boldsymbol{\theta}\mathbf{,y}_{1:T})$ with respect to the relevant posterior
approximation. The methods in (iii) have been based on the principles of
approximate Bayesian computation (ABC)
(\citealp{marin:pudlo:robert:ryder:2011}; \citealp{sisson2011likelihood};
\citealp{sisson2018handbook}), Bayesian synthetic likelihood (BSL)
(\citealp{price2018bayesian}), variational Bayes (VB)
(\citealp{blei2017variational}), and integrated nested Laplace approximation
(INLA) (\citealp{rue:martino:chopin:2009}), and produce what are termed
`approximate' forecast, or predictive distributions. Suffice to say that the
principle adopted for estimating the `approximate predictive' so defined is
typically one and the same: draws of $\boldsymbol{\theta}$ from the
approximate posterior (however produced) are used to produce either a sample
mean of conditional predictives, or $M$ draws of $y_{T+1}$ from $p(y_{T+1}%
|\boldsymbol{\theta}\mathbf{,y}_{1:T})$, with kernel density estimation then applied.

The production of (\ref{mod_av}) requires the computation of each
model-specific predictive, plus the computation of each (\ref{marg_k}). The
first set of $K$ computations would proceed via the sorts of steps outlined
above. Computation of the $K$ marginal likelihoods could also be performed via
one of the three broad methods listed above (in particular (ii)\textit{ }or
(iii)); however, the fact that each (\ref{marg_k}) is a prior, rather than a
posterior expectation does have implications for precise manner in which
computation is implemented. (See \citealp{ARDIA2012}, and
\citealp{llorente2021marginal}, for details).

\section{Bayesian Forecasting: A Chronological Tour\label{hist_persp}}

\subsection{The late 20th century: The advent of MCMC\label{sol}}


As is clear from the brief synopsis above, it is \textit{simulation }that is
key to computing forecast distributions when they are not available in closed
form. While the use of simulation to compute statistical quantities of
interest was known by the 1970s (\citealp{metropolis:ulam:1949};
\citealp{metropolis:1953}; \citealp{hammersley:handscomb:1964};
\citealp{hastings:1970}), the technology required to perform simulation in a
convenient and timely fashion was not yet available, and simulation-based
computation thus remained largely out of reach. To quote
\cite{geweke2006bayesian}:\bigskip

\begin{center}
\textit{\textquotedblleft In the beginning, there was diffuseness, conjugacy
and analytical work!\textquotedblright}

\bigskip
\end{center}

In the latter part of the 20th century, things changed. The increased speed
and availability of desktop machines (\citealp{ceruzzi:2003}), allied with
critical advances in simulation methodology, led to\textbf{ }a proliferation
of methods for accessing $p(y_{T+1}|\mathbf{y}_{1:T})$ via the simulation of
draws from $p(\boldsymbol{\theta}\mathbf{|y}_{1:T})$. To this end, we give a
brief outline of the pre-eminent posterior simulation algorithms of the 1990s
(and the early 2000s): Gibbs sampling (Section \ref{Gibbs_sampl}),
MH-within-Gibbs sampling (Section \ref{mh_within_Gibbs}), and (MH-within-)
Gibbs sampling allied with data augmentation (Section \ref{aug}); touching on
the types of forecasting models that were able to be treated via such methods,
most notably the ubiquitous state space models that underpin much modern
Bayesian forecasting. To keep the exposition concise, we place all algorithmic
details in\textbf{ }Appendix \ref{app:algos}, and reference specific
algorithms from Appendix \ref{app:algos} at suitable points in the text.

\subsubsection{Gibbs sampling\label{Gibbs_sampl}}

As a general rule, if $p(\boldsymbol{\theta}\mathbf{|y}_{1:T})$ does not have
a closed-form representation, it is also not amenable to Monte Carlo sampling,
as the latter requires that $p(\boldsymbol{\theta}\mathbf{|y}_{1:T})$ can be
decomposed into recognizable densities, from which computer simulation is
feasible. IS (\citealp{kloek1978bayesian}; \citealp{geweke:1989}), via use of
an `importance' or `proposal' density, $q(\boldsymbol{\theta}\mathbf{|y}%
_{1:T})$, that matches $p(\boldsymbol{\theta}\mathbf{|y}_{1:T})$ well and
which \textit{can }be drawn from, is a possible solution in some cases.
However, the algorithm can fail to produce representative draws from
$p(\boldsymbol{\theta}\mathbf{|y}_{1:T})$ when the dimension of
$\boldsymbol{\theta}$ is large, due to the difficulty of finding a
$q(\boldsymbol{\theta}\mathbf{|y}_{1:T})$ that is a `good match' to
$p(\boldsymbol{\theta}\mathbf{|y}_{1:T})$ in high dimensions.

In contrast, under certain conditions, a Gibbs sampler is able to produce a
(dependent) set of draws from the \textit{joint }posterior via iterative
sampling from lower dimensional, and often standard, \textit{conditional
}posteriors. In other words, a Gibbs sampler takes advantage of the fact that,
while joint and marginal posterior distributions are usually complex in form
and unable to be simulated from directly, conditional posteriors are often
standard and amenable to simulation. Given the satisfaction of the required
convergence conditions (\citealp{geyer2011introduction}), draws\textbf{
}$\boldsymbol{\theta}^{(i)}$, $i=1,2,...,M$, produced via iterative sampling
from the full conditionals converge in distribution to $p(\boldsymbol{\theta
}\mathbf{|y}_{1:T})$ as $M\rightarrow\infty$, and can be used to produce
a\textbf{ }$\sqrt{M}$-consistent estimate of the ordinates of\textbf{
}$p(y_{T+1}|\mathbf{y}_{1:T})$\textbf{ }across the support of\textbf{
}$y_{T+1}$\textbf{ }in the manner described in Section \ref{comp}. Decisions
about how to partition, or `block'\textbf{ }$\boldsymbol{\theta}$\textbf{
}need to be made (\citealp{liu:won:kon94}; \citealp{roberts:sahu:1997}), with
a view to increasing the `efficiency' of the chain which, in effect, amounts
to ensuring an accurate estimate of\textbf{ }$p(y_{T+1}|\mathbf{y}_{1:T})$ for
a given number of draws,\textbf{ }$M.$ (See Algorithm \ref{Gibbsalg1} in
Appendix \ref{A1}.)

\cite{CHIB1993} and \cite{McCulloch1994} are the earliest examples of using
Gibbs algorithms for Bayesian estimation and prediction in time series
settings; both papers exploiting the fact that despite\textbf{ }%
$p(\boldsymbol{\theta}\mathbf{|y}_{1:T})$ and $p(y_{T+1}|\mathbf{y}_{1:T})$
precluding analytical treatment in most of the examples considered, the
conditional posteriors always have closed forms. As one would anticipate
however, a `pure' Gibbs algorithm based on a full set of standard conditionals
is not always possible, with the more typical situation being one in which one
or more of the conditionals -- associated with any given partitioning of the
parameter space -- are not available in closed form. The following section
describes how to adapt a Gibbs algorithm in cases where certain conditional
components are not known in closed form, and, in so doing, illustrates a
powerful simulation-based\textbf{ }algorithm for accessing\textbf{ }%
$p(y_{T+1}|\mathbf{y}_{1:T})$ in more complex settings.

\subsubsection{MH-within-Gibbs sampling\label{mh_within_Gibbs}}

The Gibbs sampler is only one example of an MCMC algorithm. The first such
example -- the `Metropolis' algorithm -- appeared in a paper that has assumed
an important status in the history of statistics: \cite{metropolis:1953}%
\footnote{For example, \cite{dongarra2000guest} rank the Metropolis algorithm
proposed in \citeauthor{metropolis:1953} as one of the 10 algorithms
\textquotedblleft with the greatest influence on the development and practice
of science and engineering in the 20th century\textquotedblright.}. The
Metropolis algorithm was subsequently generalized by \cite{hastings:1970}, and
it is this `MH' version of the method that is typically referenced. For the
purpose of this review, the key role of the MH algorithm is to enable sampling
from non-standard conditionals within a Gibbs algorithm, in particular when
the dimension of the conditionals precludes (say) the exclusive use of inverse
cumulative distribution function\textbf{ (}ICDF) sampling.\footnote{Any
non-standard probability distribution can, in principle, be drawn from using
ICDF sampling. The term `Griddy Gibbs' sampling was first used by
\cite{ritter:tanner:1992} to refer to the use of ICDF sampling to draw from
non-standard conditionals in a Gibbs scheme. Given that the method amounts to
the use of numerical quadrature, it suffers from the curse of dimensionality,
and is thus infeasible for drawing from anything other than very
low-dimensional conditionals. See \cite{Bauwens1998}\textbf{ }for the
application of the Griddy-Gibbs sampler to a generalized autoregressive
conditionally heteroscedastic (GARCH) model for financial returns.}

Under regularity, a Markov chain that converges to $p(\boldsymbol{\theta
}\mathbf{|y}_{1:T})$ can be produced by embedding an MH algorithm (or MH
\textit{algorithms}) within an outer Gibbs loop. In short, an MH-within-Gibbs
algorithm proceeds by drawing from any non-standard conditional indirectly,
via a `candidate', or `proposal' distribution that is deemed to be a good
match to the inaccessible conditional, and accepting the draw with a given
probability. Critically, the formula that defines the acceptance probability
involves evaluation of the non-standard conditional only up to its integrating
constant; hence the conditional need not be known in its
entirety.\footnote{Moreover, and in contrast to IS, the requirement to find a
well-matched proposal distribution is facilitated by the dimension reduction
invoked by the breaking down of the high-dimensional joint posterior into the
lower dimensional conditionals, before any proposal distribution needs to be
specified.} Again, under appropriate regularity, the draws $\boldsymbol{\theta
}^{(i)}$, $i=1,2,...,M$, from the MH-within-Gibbs algorithm\textbf{ }converge
in distribution to $p(\boldsymbol{\theta}\mathbf{|y}_{1:T})$ as $M\rightarrow
\infty$, and can be used to produce a $\sqrt{M}$-consistent estimate of the
ordinates of $p(y_{T+1}|\mathbf{y}_{1:T})$. (See Algorithm
\ref{MHwithinGibbsalg1} in Appendix \ref{A2}.)

As will become evident in the subsequent empirical review sections,
MH-within-Gibbs algorithms remain the dominant form of method used to sample
from posteriors -- and to estimate predictive distributions -- for time series
models for which a convenient partitioning of the parameter space is
available, and for which the conditional posteriors are known up to their
integrating constants. Hence, we reserve further elaboration on the use of
such algorithms in practice until the appropriate points in Section
\ref{reviews}.

\subsubsection{ MCMC, data augmentation, and state space models\label{aug}}

For many empirical problems in economics and related fields, a suitable model
can be partitioned into two sets: static unknowns $\boldsymbol{\theta}$, which
are fixed throughout time, and latent data, $\mathbf{z}_{1:T}=(z_{1}%
,z_{2},...,z_{T})^{\prime}$, which vary over time. The latent states may be
intrinsic to the model -- as in a state space model -- or may be auxiliary
variables introduced purely for the purpose of facilitating posterior
sampling. Application of a Gibbs-based MCMC scheme to the joint, or
`augmented' set of unknowns $\left(  \boldsymbol{\theta},\mathbf{z}%
_{1:T}\right)  $ is often referred to as `data augmentation', in the spirit of
\cite{tanner87}, and such schemes have enabled the Bayesian analysis of large
classes of time series models that would otherwise have been inaccessible.

We illustrate here the basic principles of the approach using a state space
model governed by a measurement density for the observed scalar random
variable, $y_{t}$, and a Markov transition density for a scalar state
variable, $z_{t}$,%
\begin{align}
&  p(y_{t}|z_{t},\boldsymbol{\theta})\label{meas}\\
&  p(z_{t}|z_{t-1},\boldsymbol{\theta}). \label{state}%
\end{align}
Using the generic notation in (\ref{meas}) and (\ref{state}), the augmented
posterior is%
\begin{equation}
p(\boldsymbol{\theta},\mathbf{z}_{1:T}|\mathbf{y}_{1:T})\propto p(\mathbf{y}%
_{1:T}|\mathbf{z}_{1:T},\boldsymbol{\theta})p(\mathbf{z}_{1:T}%
|\boldsymbol{\theta})p(\boldsymbol{\theta}). \label{aug_post}%
\end{equation}
In certain cases, the model structure is such that a pure Gibbs scheme can be
used to produce draws from $p(\boldsymbol{\theta},\mathbf{z}_{1:T}%
|\mathbf{y}_{1:T})$ and, thus, from $p(\boldsymbol{\theta}|\mathbf{y}_{1:T})$;
an insight obtained independently by \cite{carter:kohn:1994} and
\cite{fruhwirth-schnatter:1994} for the case of the linear Gaussian
state-space model, for example. However, implementation of such a scheme will,
by definition, require both\textbf{ }$p(\boldsymbol{\theta}|\mathbf{z}%
_{1:T},\mathbf{y}_{1:T})$ and\textbf{ }$p(\mathbf{z}_{1:T}|\boldsymbol{\theta
},\mathbf{y}_{1:T})$ to have recognizable forms. In more general cases, in
which either the measurement or state\textbf{ }equation has non-linear and/or
non-Gaussian features, the resulting conditionals will not necessarily have a
known closed form, which necessitates the addition of MH steps within the
outer Gibbs loop. Such a treatment was the method of attack for large classes
of models in the 1990s and early 2000s. Relevant contributions here, which
include specific treatments of the ubiquitous stochastic volatility (SV)
model, are \cite{carlin:polson:stoffer:1992}, \cite{jacquier94},
\cite{shepard97}, \cite{kim1998svl}, \cite{chib:2002}, \cite{stroud2003},
\cite{CHIB2006}, \cite{STRICKLAND2006}, \cite{ChibOMORI2007} and
\cite{STRICKLAND2008}. The reviews of \cite{fearnhead2011mcmc} and
\cite{giordani2011bayesian} provide more detailed accounts and extensive
referencing of this earlier literature.\footnote{We also note here the work of
\cite{chib94}, in which the state space representation of an autoregressive
moving average (ARMA(p,q)) model (\citealp{Harvey1981}) was exploited, and the
principle of data augmentation invoked, in order to enable an MH-within-Gibbs
scheme to be applied.} (See also Appendix \ref{A3}.)

To conclude, and once again using the generic notation in (\ref{meas}) and
(\ref{state}), once draws have been produced from $p(\boldsymbol{\theta
},\mathbf{z}_{1:T}|\mathbf{y}_{1:T})$, the predictive pdf,%
\begin{equation}
p(y_{T+1}|\mathbf{y}_{1:T})=\int_{{z}_{T+1}}\int_{\mathbf{z}_{1:T}}%
\int_{{\Theta}}p(y_{T+1}|z_{T+1},\boldsymbol{\theta}\mathbf{,y}_{1:T}%
)p(z_{T+1}|z_{T},\boldsymbol{\theta})p(\boldsymbol{\theta},\mathbf{z}%
_{1:T}|\mathbf{y}_{1:T})d\boldsymbol{\theta}d\mathbf{z}_{1:T}d{z}_{T+1},
\label{forecast_ss}%
\end{equation}
can be estimated in the usual way, using subsequent draws from $p(z_{T+1}%
|z_{T},\boldsymbol{\theta})$ and $p(y_{T+1}|z_{T+1},\boldsymbol{\theta
}\mathbf{,y}_{1:T})$, or by averaging the conditional predictives over all
draws of $z_{T+1}$ and $\boldsymbol{\theta}$.

\subsection{The 21st Century: Intractable forecasting models\label{intrac}}

\subsubsection{What do we mean by `intractable'?}

The MCMC methods that evolved during the late 20th century continue to serve
as the `bread and butter' of Bayesian forecasting, as will be made evident in
Section \ref{reviews}. Nevertheless, more ambitious forecasting problems are
now being tackled, and this has tested the mettle of some of the early
algorithms. As a consequence, Bayesian forecasters have begun to exploit more
modern computational techniques, and it is those techniques that we touch on
briefly in this section.

It is convenient to characterize these newer computational developments as
different types of solutions to so-called `intractable' forecasting problems,
by which we mean: (a) forecasts based on models with data generating processes
(DGPs) that cannot be readily expressed as a pdf, or probability mass function
(pmf); (b)\textit{ }forecasts based on high-dimensional models, with a very
large number of unknowns; (c)\textit{ }forecasts produced using extremely
large data sets. Problems that feature problem (a)\textit{ }are referred to
as\textit{ doubly-intractable }problems, as not only is $p(\boldsymbol{\theta
}|\mathbf{y}_{1:T})$ not available in its entirety (as is typical), but the
DGP itself is also not able to be expressed analytically.

With reference to (a), the MCMC methods referenced so far entail the
evaluation of the DGP as a pd(/m)f, either in the calculation of the
acceptance probability in any MH sub-step, or in the specification of full
conditionals in any `pure' Gibbs step. Hence, they are infeasible when DGPs do
not admit such a representation. Many such DGPs exist (see, for example,
\citealp{martin2021approximating}, for a list of examples); however,
particularly pertinent ones to mention here are continuous time models in
finance with unknown transition densities \citep{gallant1996moments}, $\alpha
$-stable models for financial returns (and/or their volatility)
(\citealp{peters2012likelihood}; {\citealp{martin2019auxiliary}}), and
stochastic dynamic equilibrium models in economics
(\citealp{calvet2015accurate}). With regard to (b), whilst, in principle (and
under appropriate regularity), a convergent MCMC chain can be constructed for
any model, the exploration of a very high-dimensional parameter space via an
MCMC algorithm \textit{can }be prohibitively slow
(\citealp{tavare:balding:griffith:donnelly:1997};
{\citealp{rue:martino:chopin:2009}}; \citealp{braun2010variational};
\citealp{lintusaari2017fundamentals}; \citealp{betancourt:2018};
\citealp{johndrow2019mcmc}). Hence, in models with a very large number of
unknowns -- including those with multiple sets of latent variables -- the
production of an accurate MCMC-based estimate of $p(y_{T+1}|\mathbf{y}_{1:T})$
in a practical amount of time may not be possible. Finally, regarding
(c)\textit{ }MCMC schemes require pointwise (i.e. for each $y_{t}$) evaluation
of $p(\mathbf{y}_{1:T}|\boldsymbol{\theta})$ at each draw of
$\boldsymbol{\theta}$, thereby inducing an $O(T)$ computational burden at each
iteration in an MCMC chain.\footnote{We recall that a sequence $X_{T}$ is
$O(T)$ if $|X_{T}/T|$ is bounded as $T\rightarrow+\infty$.} Such schemes can
thus struggle when confronted with `big data' \citep{bardenet2017markov}. In
this context, `big data' refers to situations where, due to the length and/or
size of the data set, the repeated evaluation of the likelihood function that
is required to produce draws from the corresponding MCMC chain is too time
consuming for the algorithm to run in a reasonable amount of time.

The methods in the following sections have been designed to solve one or more
of these instances of intractability. The techniques in Section \ref{exact} do
so whilst preserving the `exact' nature of the estimate of $p(y_{T+1}%
|\mathbf{y}_{1:T})$, whilst those in Section \ref{approx} aim to produce an
approximation of $p(y_{T+1}|\mathbf{y}_{1:T})$ only.

\subsubsection{Exact computational solutions\label{exact}}

The first two decades of the 21st century have witnessed a wealth of advances
in both\textbf{ }MCMC and IS-based algorithms. The goal of the newer MCMC
algorithms -- at their heart -- is to explore the high mass region of the
joint posterior more \textit{efficiently}, in particular when the dimension of
the space of unknowns is large. This, in turn, enables a more accurate
estimate of $p(y_{T+1}|\mathbf{y}_{1:T})$ to be produced for a given
computational budget. This goal has been achieved via a variety of means,
which (in the spirit of \citealp{robert2018accelerating}, and
\citealp{martin2022history})\textbf{ }can be summarized as follows: the use of
more geometric information about the target posterior, most notably the use of
Hamiltonian updates (\citealp{neal:2011}; \citealp{hoffman2014no}); the use of
better MH candidate, or {proposal distributions, including those that `adapt'
to previous draws (\citealp{nott2005adaptive}; \citealp{roberts2009examples})}%
; various types of combinations of multiple chains
(\citealp{jacob:robert:smith:2010}; \citealp{neal2011mcmc};
\citealp{neiswanger:wang:xing:2013}; \citealp{glynn:rhee:2014};
{\citealp{huber2016perfect}}; \citealp{jacob2019unbiased}); or {the use of
ex-post variance reduction methods (\citealp{craiu:meng:2005};
\citealp{douc:robert:2011}; \citealp{owen2017statistically};
\citealp{baker2019}). We refer the reader to} \cite{greenetal2015},
\cite{robert2018accelerating} and \cite{dunson2019hastings} for detailed
reviews of modern developments in MCMC, and to \cite{Jahan2020} for an
overview of the way in which certain of the newer methods manage the problem
of scale -- in terms of either the unknowns or the data, or both.

Whilst not designed expressly to deal with problems of scale, sequential Monte
Carlo (SMC) methods -- which exploit the principles of IS -- have developed in
parallel to the expansion of the MCMC stable. Devised initially for the
sequential analysis of state space models, via methods of `particle filtering'
\citep{gordon:salmon:smith:1993}, SMC methods have evolved into a larger suite
of methods used to perform both sequential and non-sequential tasks
(\citealp{naesseth2019elements}; \citealp{Chopin2020}). For the purpose of
this review, the most pertinent development is the melding of particle
filtering with MCMC in state space settings to produce a particle marginal MH
(PMMH) algorithm (\citealp{andrieu:doucet:holenstein:2010};
\citealp{flury_shephard_2011}; \citealp{pitt2012some};
\citealp{doucet2015efficient}; \citealp{deligiannidis2018correlated}). Such
algorithms tackle intractability type (a) in the dichotomy of the previous
section, by replacing an `unavailable' likelihood function by an unbiased
estimate -- produced via the particle filter -- in an MH algorithm which,
under regularity, retains the posterior $p(\boldsymbol{\theta}|\mathbf{y}%
_{1:T})$ as its invariant distribution. Given the increasingly important role
played by PMMH, a brief algorithmic description of it is included in Algorithm
\ref{PMMHalg1} in Appendix \ref{A4}.\footnote{PMMH is actually a special case
of the general pseudo-marginal MH technique (also sometimes denoted by the
abbreviation `PMMH'), in which a pseudo likelihood, produced -- in some manner
or another -- as an unbiased estimator of the true likelihood, is used within
an MH algorithm. See, for example, the subsampling methods based on
pseudo-marginal MCMC (\citealp{bardenet2017markov};
\citealp{quiroz2018speeding}; \citealp{quiroz2019speeding}) used expressly to
improve the performance of MCMC in the case of a large-dimensional
$\mathbf{y}_{1:T}$ (i.e. intractability type (c)\textit{).}}

\subsubsection{Approximate computational solutions\label{approx}}

In situations in which the dimension, or structure of the forecasting model,
or the size of the data set, still precludes the use of either an MCMC or a
PMMH approach, an approximate method may be the only computational
option.\textbf{ }The cost of adopting such a solution is that these methods no
longer directly target the exact predictive, $p(y_{T+1}|\mathbf{y}_{1:T}%
)$;\textbf{ }instead, an approximation of $p(y_{T+1}|\mathbf{y}_{1:T}%
)$\textbf{ }becomes the goal.

The spirit of these methods is to approximate $p(y_{T+1}|\mathbf{y}_{1:T})$
via some feasible approximation to the posterior $p(\boldsymbol{\theta
}|\mathbf{y}_{1:T})$. Denoting the posterior approximation generically by
$g(\boldsymbol{\theta}|\mathbf{y}_{1:T})$, the resultant approximate
predictive can be expressed as%
\begin{equation}
g(y_{T+1}|\mathbf{y}_{1:T})=\int_{{\Theta}}p(y_{T+1}|\boldsymbol{\theta
}\mathbf{,y}_{1:T})g(\boldsymbol{\theta}|\mathbf{y}_{1:T})d\boldsymbol{\theta
}, \label{approx_pred}%
\end{equation}
in the case where there are only static unknowns. When the model features both
static parameters and time-varying latent parameters, and exploiting the
Markov property of the state process in (\ref{state}), the approximate
predictive can be represented as%
\begin{equation}
g(y_{T+1}|\mathbf{y}_{1:T})=\int_{{z}_{T+1}}\int_{z_{T}}\int_{{\Theta}%
}p(y_{T+1}|z_{T+1},\boldsymbol{\theta}\mathbf{,y}_{1:T})p(z_{T+1}%
|z_{T},\boldsymbol{\theta})p(z_{T}|\boldsymbol{\theta},\mathbf{y}%
_{1:T})g(\boldsymbol{\theta}|\mathbf{y}_{1:T})d\boldsymbol{\theta}dz_{T}%
d{z}_{T+1}. \label{eq12}%
\end{equation}
Given draws of\textbf{ }$\boldsymbol{\theta}$\textbf{ }from\textbf{
}$g(\boldsymbol{\theta}|\mathbf{y}_{1:T})$, and given an appropriate
forward-filtering algorithm to draw from\textbf{ }$p(z_{T}|\boldsymbol{\theta
},\mathbf{y}_{1:T})$\textbf{ }when needed, a simulation-based estimate of
$g(y_{T+1}|\mathbf{y}_{1:T})$\textbf{ }can be produced in the usual way,
either as a sample mean of the conditional predictives defined by the draws
of\textbf{ }$\boldsymbol{\theta}$\textbf{ }(and\textbf{ }$z_{T+1}$), or by
applying kernel density techniques to the draws of\textbf{ }$y_{T+1}$\textbf{
}from the conditional predictive.

With reference to the taxonomy of intractable problems delineated in Section
\ref{intrac}, the different methods of producing\textbf{ }%
$g(\boldsymbol{\theta}|\mathbf{y}_{1:T})$ (and, hence,\textbf{ }%
$g(y_{T+1}|\mathbf{y}_{1:T})$) can be categorized according to whether they
are being used\textbf{ }to obviate (a)\textit{ }or to\textbf{ }tackle a
problem of scale: (b)\textit{ }and/or (c)\textit{. }Both ABC and BSL avoid the
need to evaluate the DGP and, hence, are feasible methods in the\textbf{
}doubly-intractable\textit{ }settings of category (a). In brief, both methods
require only \textit{simulation}, not \textit{evaluation}, of the DGP. The
\textit{approximation} of $p(\boldsymbol{\theta}|\mathbf{y}_{1:T})$ arises,
primarily, from the fact that both methods -- in different ways -- degrade the
information in the full data set,\textbf{ }$\mathbf{y}_{1:T}$,\textbf{ }to the
information contained in a set of summary statistics, $\eta(\mathbf{y}_{1:T}%
)$.\textbf{ }As such,\textbf{ }the target becomes the so-called `partial'
posterior for $\boldsymbol{\theta}$,\textbf{ }which conditions on\textbf{
}$\eta(\mathbf{y}_{1:T})$, rather than\textbf{ }$\mathbf{y}_{1:T}$.\textbf{
}The quality of the approximation is thus dependent on the informativeness of
the summaries, as well as on other forms of approximation invoked in the
implementation of the methods. Vanilla versions of both algorithms are
provided in Algorithms \ref{ABCalg2} (Appendix \ref{A5}) and \ref{BSLalg1}
(Appendix \ref{A6}) respectively.

In contrast to ABC and BSL, VB and INLA still \textit{target} the exact
posterior $p(\boldsymbol{\theta}|\mathbf{y}_{1:T})$, but provide
approximations that can be computationally convenient when the \textit{scale}
of the empirical problem is large in some sense (so problem (b)\textit{
}and/or problem\textbf{ }(c)), often as a consequence of the specification of
a high number of latent, or `local', parameters in the model, in addition to
the (usually) smaller set of `global' parameters ($\boldsymbol{\theta}$ in our
notation). Adopting the technique of the calculus of variations, VB produces
an approximation of $p(\boldsymbol{\theta}\mathbf{|y}_{1:T})$ that is
`closest' to $p(\boldsymbol{\theta}\mathbf{|y}_{1:T})$ within a chosen
\textit{variational family}, whilst INLA applies a series of nested Laplace
approximations (\citealp{laplace:1774}; \citealp{tierney:kadane:1986};
\citealp{tierney:kass:kadane:1989}) to a high-dimensional latent Gaussian
model to produce an approximation of $p(\boldsymbol{\theta}|\mathbf{y}_{1:T}%
)$. Both VB and INLA exploit state-of-the-art optimization techniques, for the
purpose of minimizing the `distance' between $p(\boldsymbol{\theta}%
\mathbf{|y}_{1:T})$ and the variational approximation in the case of VB, and
for the purpose of producing the mode of the high-dimensional vector of latent
states in the case of INLA. The basic principles of VB and INLA are provided
in Appendices \ref{A7} and \ref{A8} respectively.

We refer the interested reader to \cite{martin2021approximating} for an
extensive review of all of these approximate Bayesian methods, as well as more
complete coverage of the existing literature, including references to in-depth
reviews of specific methods. \citeauthor{martin2021approximating} also
includes discussion of `hybrid' methods that mix and match features of more
than one computational technique, with the aim of tackling multiple instances
of `intractability' simultaneously.

Regardless of which approximation method is used, the hope is that the
resulting approximate predictive $g(y_{T+1}|\mathbf{y}_{1:T})$ performs well
relative to the inaccessible exact predictive, and that issue is addressed in
certain work cited in the empirical reviews in\textbf{ }Section \ref{reviews}.

\subsection{The 21st Century: Misspecified forecasting models}

\subsubsection{The role of model specification in Bayesian forecasting}

Inherent in the{ conventional Bayesian approach to forecasting is the
assumption that the process that has generated the observed data tallies with
the particular model that underpins the likelihood function. }Bayesian model
averaging (BMA) -- and the resultant predictive in (\ref{mod_av}) -- has
evolved as a principled way of catering for uncertainty about the predictive
model, and BMA remains a very important technique in the Bayesian toolbox.
Nevertheless, underpinning BMA is still the assumption that the true process{
is spanned by the set of models over which one averages -- i.e. that the
so-called }$\mathcal{M}$-closed view of the world
({\citealp{bernardo:smith:1994}}) prevails.

In response to these perceived limitations of the conventional approach,
attention has recently been given to producing predictions that are `fit for
purpose', by focusing the Bayesian machinery on the \textit{specific} goals of
the predictive analysis at hand. In the following sections we briefly
summarize three such approaches, all of which move beyond the conventional
likelihood-based Bayesian update, and $\mathcal{M}$-closed paradigm: seeking
to produce accurate predictions without recourse to the assumption of correct
model specification.

\subsubsection{Focused, or `loss-based' Bayesian prediction\label{focus}}

\cite{loaiza2019focused}{ propose {an }approach to Bayesian prediction
expressly designed for the context of misspecification. In brief, rather than
a correct predictive model being assumed, a}{ prior is placed over a class of
\textit{plausible} predictive models. The prior is then updated to a posterior
via a sample criterion function that is constructed using a{ \textit{scoring
rule }}}({\citealp{gneiting2007strictly}}) that rewards the type of predictive
accuracy (e.g. accurate prediction of extreme values) that is important for
the particular empirical problem being tackled{{. With}} a criterion function
that explicitly captures predictive accuracy{{ replacing the likelihood
function in the Bayesian update, the explicit need for correct model
specification is avoided.}}

Following \cite{gneiting2007strictly}, and using generic notation,\ for
$\mathcal{P}$ a convex class of predictive distributions on $(\Omega
,\mathcal{F})$, the predictive accuracy of $P\in\mathcal{P}$ can be assessed
using a scoring rule $S:\mathcal{P}\times\Omega\rightarrow\mathbb{R}$. If the
value $y$ eventuates, then the positively-oriented `score' of the predictive
$P$, is $S(P,y).$ The expected score under the true unknown predictive $P_{0}$
{is }defined as
\begin{equation}
\mathbb{S}(\cdot,P_{0}):=\int_{y\in\Omega}S(\cdot,y)dP_{0}(y).\label{score}%
\end{equation}
A scoring rule is said to be proper relative to $\mathcal{P}$ if, for all
$P,G\in\mathcal{P}$, $S(G,G)\geq S(P,G),$ and is strictly proper, relative to
$\mathcal{P}$, if $S(G,G)=S(P,G)\iff P=G$. Scoring rules are important
mechanisms as they elicit truth telling within the forecasting exercise: if
the true predictive $P_{0}$ \textit{were} known, then in terms of forecasting
accuracy as measured by the scoring rule $S(\cdot,\cdot)$ it would be optimal
to use $P_{0}$.

Different scoring rules rewards different forms of predictive accuracy (see
{\citealp{gneiting2007strictly}}, {\citealp{Opschoor2017}, and
\citealp{martin2022optimal}} for expositions); hence the motivation to drive
the update by the score that `matters'. Since $P_{0}$ and the expected score
$\mathbb{S}(\cdot,P_{0})$ are unattainable in practice, an estimate based on
$\mathbf{y}_{1:T}$ is used to define the sample criterion, $S_{T}%
(\boldsymbol{\theta}):=\sum_{t=0}^{T-1}S[p(y_{t+1}|\boldsymbol{\theta
},\mathbf{y}_{1:t}),y_{t+1}]$, where $p(y_{t+1}|\boldsymbol{\theta}%
,\mathbf{y}_{1:t})$ is the pdf associated with a given $P.$ Adopting the
exponential updating rule proposed by \cite{bissiri:etal:2016} (see also
\citealp{giummole2017objective}, \citealp{holmes2017assigning},
\citealp{guedj2019}, \citealp{lyddon2019general}, and
\citealp{syring2019calibrating}), \cite{loaiza2019focused} define the
\textit{generalized }(or \textit{Gibbs}) posterior:
\begin{equation}
\pi_{w}(\boldsymbol{\theta}|\mathbf{y}_{1:T})=\frac{\exp\left[  wS_{T}%
(\boldsymbol{\theta})\right]  \pi(\boldsymbol{\theta})}{\int_{\Theta}%
\exp\left[  wS_{T}(\boldsymbol{\theta})\right]  \pi(\boldsymbol{\theta
})d\boldsymbol{\theta}},\label{post}%
\end{equation}
for some learning rate $\omega\geq0$, calibrated in a preliminary step. This
posterior explicitly places high weight on -- or \textit{focuses} on
--\textbf{ }values of $\boldsymbol{\theta}$ that yield high predictive
accuracy in the scoring rule $S(\cdot,\cdot)$. As such, the process of
building a Bayesian predictive as:
\begin{equation}
p_{w}(y_{T+1}|\mathbf{y}_{1:T})=\int_{\Theta}p(y_{T+1}|\boldsymbol{\theta
}\mathbf{,y}_{1:T})\pi_{w}(\boldsymbol{\theta}|\mathbf{y}_{1:T}%
)d\boldsymbol{\theta},\label{Gibbs}%
\end{equation}
is termed `focused Bayesian prediction'\textit{ }(FBP) by the authors. By
construction, when the predictive model, $p(y_{T+1}|\boldsymbol{\theta
}\mathbf{,y}_{1:T})$, is misspecified, (\ref{Gibbs}) will -- out-of-sample --
often outperform, in the chosen rule $S(\cdot,\cdot)$, the likelihood (or
{log-score})-based predictive in (\ref{exact_pred}), and this is demonstrated
in \citeauthor{loaiza2019focused} both theoretically and in extensive
numerical illustrations.

Since a positively-oriented score can, equivalently, be viewed as the negative
of a measure of predictive loss, FBP can also be referred to as `loss-based'
prediction. Such terminology is indeed adopted in {\cite{frazierloss}, in
which the principles delineated here are extended to high-dimensional models,
and approximations to both }$\pi_{w}(\boldsymbol{\theta}|\mathbf{y}_{1:T})$
and $p_{w}(y_{T+1}|\mathbf{y}_{1:T})$ based on VB proposed, and validated. We
note that the term `loss'\ as it is used in \cite{loaiza2019focused} and
\cite{frazierloss} refers specifically to predictive loss as quantified by a
proper scoring rule. For the application of loss-based Bayesian
\textit{inference}, in which more general forms of loss functions may drive
the Bayesian update, we refer the reader to certain of the other literature
cited above, namely \cite{bissiri:etal:2016}, \cite{holmes2017assigning},
\cite{lyddon2019general} and \cite{syring2019calibrating}.

\subsubsection{Bayesian predictive combinations: Beyond BMA\label{comb}}

The predictive distributions within the `plausible class' referenced above
{may characterize a single dynamic structure} depending on a vector of unknown
parameters, $\boldsymbol{\theta}$, {or may constitute} weighted combinations
of predictives from distinct {models}, in which case $\boldsymbol{\theta}$
comprises both the model-specific parameters and the weights. {As such, FBP
provides a coherent Bayesian }method for{\ estimating weighted combinations of
predictives via predictive accuracy criteria, and without the need to assume
that the true model is spanned by the set of constituent predictives -- an
assumption that underpins BMA, as we have noted. }

{A similar motivation underlies other contributions to the extensive Bayesian
literature on estimating combinations of predictives that has now developed
(and which rivals the large }frequentist literature on forecast combinations
that has also evolved\footnote{See {\cite{Hall2007}, \cite{ranjan2010},
\cite{Geweke2011} and \cite{gneiting:2013}} for early contributions to the
frequentist forecast combination literature, and \cite{Wang2022} for a recent
review. We note that whilst {\cite{Geweke2011} is not explicitly Bayesian, in
terms of estimating the optimal predictive combination, it provides important
insights into the connection between the `optimal linear pool' and BMA, and
also uses Bayesian numerical methods in the production of some of the
constituent forecast distributions.}}), {with predictive performance} --
quantified by a range of user-specified measures of predictive accuracy --
{driving the posterior updating of the weights. }Indeed, {the Bayesian}
literature,{\ having access as it does to powerful computational tools, has
been able to invoke more complex weighting schemes than can be tackled via
frequentist (optimization) methods. Notable contributions}, including some
also driven by the criterion of predictive calibration (\citealp{Dawid1982};
\citealp{Dawid1985}; \citealp{gneiting2007probabilistic}), {include
}{\cite{Billio2013}}, {{\cite{casarin2015jss}}, \cite{casarin2015},
{\cite{casarin2016}}, {\cite{Pett2016}}}, \cite{aastveit2018},
{{{\cite{Bassetti2018}, \cite{BASTURK2019}}} and {\cite{CASARIN2023}. Once
again adopting the language of \cite{bernardo:smith:1994}}}, this literature
seeks to move Bayesian predictive combinations beyond the $\mathcal{M}$-closed
world of {BMA to the }$\mathcal{M}$-open world that accords with the reality
of misspecification.

We complete this section by also highlighting one particular generalization of
BMA that aims, not so much to cater for the $\mathcal{M}$-open world but,
rather, to remove the fixed-weight restriction that is inherent to BMA.
Certain of the references cited above either explicitly allow for the weights
attached to the constituent forecasts to evolve over the time period,
$t=1,2,....,T$, on which the predictive distribution for $y_{T+1}$ conditions
(e.g. \citealp{Billio2013}, and \citealp{CASARIN2023}) or implicitly allow for
such a possibility (e.g. \citealp{loaiza2019focused}). However, so-called
dynamic model averaging (DMA) accommodates time-varying weights via a more
direct generalization of BMA, and nests BMA when appropriate settings are
activated (see \citealp{Koop2012}, page 875 for an illustration of this). We
refer the reader to \cite{raftery:2010} for the initial proposal of DMA,
\cite{Koop2012} for the application of the method to forecasting inflation,
and \cite{Nonejad:2021} for a recent review of the methodology, with a focus
on applications in economics and finance.\footnote{We also refer the reader to
\cite{green:1995}, \citealp{madigan:raftery:1995}, and \cite{george:2000}, for
alternative approaches to catering for model uncertainty in the Bayesian
framework, In brief, such approaches -- in one way or another -- design MCMC
samplers to tackle an augmented space in which model uncertainty is
incorporated. As a consequence, the computation of any expectation of
interest, including that which defines a predictive distribution,
automatically factors in all uncertainty associated with both the parameters
of each model and the model structure itself. See \cite{green03},
\cite{marin:mengersen:robert:2005}, \cite{chib2011introduction} and
\cite{fan2011reversible} for reviews and more complete referencing.}

\subsubsection{Bayesian predictive decision synthesis}

A third approach that seeks to produce Bayesian predictions without relying
explicitly\textbf{ }on correct model specification is Bayesian predictive
synthesis (BPS) (\citealp{johnson2017bayesian}; \citealp{mcalinn2019dynamic};
\citealp{mcalinn2020multivariate}; \citealp{Aastveit2022}), recently expanded
to Bayesian predictive \textit{decision }synthesis (BPDS) by
\cite{tallman2022bayesian}. In particular, BPDS provides a sound
decision-theoretic framework for constructing forecast combinations, and can
be shown to encompass several commonly-suggested Bayesian forecasting approaches.

The starting point of BPDS is the production of a prior distribution over the
$m$-dimensional unknown outcome $\mathbf{y}$ -- implicitly indexed by $T+1$ in
a time series forecasting application --\textbf{ }and the information set
$\mathcal{H}$, encoded via the $J$ predictive models $\{h_{j}(\mathbf{y}%
|\mathbf{x}_{j}):1\leq j\leq J\}$, where $\mathbf{x}=(\mathbf{x}_{1}%
,\dots,\mathbf{x}_{J})$ denotes the collection of vectors of (possibly latent)
dummy variables associated with a decision. The decision maker then constructs
a predictive by integrating out $\mathbf{x}$ using a `synthesis function'
$\alpha(\mathbf{y}|\mathbf{x})$:
\[
p(\mathbf{y}|\mathcal{H})=\int_{\mathcal{X}}\alpha(\mathbf{y}|\mathbf{x}%
)\prod_{j=1}^{J}h_{j}(\mathbf{y}|\mathbf{x}_{j})d\mathbf{x}_{1}\dots
d\mathbf{x}_{J}.
\]
The choice of the synthesis function $\alpha(\cdot|\mathbf{x})$ can be used to
drive the analysis. For instance, in the case of forecast combinations, we can
take $h_{j}(\mathbf{y}|\mathbf{x})=p_{j}\left(  \mathbf{y}|\mathbf{x}%
,\mathcal{M}_{j}\right)  $, for some model $\mathcal{M}_{j}$, and then any set
of synthesis functions $\alpha_{j}(\cdot|\mathbf{x})$ such that the
combination density
\[
p(\mathbf{y}|\mathcal{H})=\int_{\mathcal{X}}\frac{\sum_{j=1}^{J}\omega
_{j}\alpha_{j}\left(  \mathbf{y},\mathbf{x}_{j}\mid\mathbf{x}\right)
p_{j}\left(  \mathbf{y}\mid\mathbf{x},\mathcal{M}_{j}\right)  }{\sum_{k=1}%
^{J}\omega_{k}\alpha_{k}\left(  \mathbf{y},\mathbf{x}_{k}\mid\mathbf{x}%
\right)  }d\mathbf{x}_{1}\dots d\mathbf{x}_{J}.
\]
is a valid density, for given weights $0<\omega_{j}<1,$ $\sum_{j=1}^{J}%
\omega_{j}=1.$ Specific choices of $\alpha_{j}\left(  \mathbf{y}%
,\mathbf{x}_{j}\mid\mathbf{x}\right)  $ then produce different forecast
combination methods (see, \citealp{johnson2017bayesian}, for a discussion);
for example, in the case of \cite{mcalinn2019dynamic} and
\cite{mcalinn2020multivariate}, the synthesis function is taken to be the
density of a (possibly multivariate) dynamic linear factor model.

In an attempt to `focus' the BPDS approach towards decisions that are tailored
to a specific user-chosen loss function underlying the analysis or decision at
hand, \cite{tallman2022bayesian} propose taking as their synthesis function,
$\alpha_{j}(\mathbf{y},\mathbf{x}_{j}|\mathbf{x})=\exp\{\tau^{\prime
}(\mathbf{x})S_{j}(\mathbf{y},\mathbf{x}_{j})\}$, where the score
$S_{j}(\mathbf{y},\mathbf{x}_{j})$ is a $k$-dimensional vector that measures
the utility one receives from realizing outcome $\mathbf{y}$ under decision
$\mathbf{x}_{j}$, and $\tau(\mathbf{x})$ is a vector that weights the
directional relevance of $S_{j}$\textbf{$(\mathbf{y},\mathbf{x}_{j})$}.

While the BPS framework, as a whole, can set the tenor of the predictions
towards dynamic forecast updates that produce predictions tailored to a loss
function of interest, via the choice of synthesis function $\alpha(\cdot
|\cdot)$, BPS \textit{is} ultimately tied to a `likelihood-type'
framework,\textbf{ }or at least a log-loss function, due to the presence of
the latent variables $\mathbf{x}$, which must be integrated out via assumed
predictive models, $p_{j}\left(  \mathbf{y}|\mathbf{x},\mathcal{M}_{j}\right)
$, and with these individual predictives produced using likelihood-based
Bayesian methods. While the BPDS approach can somewhat circumvent the reliance
on the likelihood, due to its ability to focus on specific scores, this
approach appears to be distinct from methods that entirely replace the
likelihood function in the update. Therefore, a very interesting research path
would involve combining the methods based on generalized posteriors discussed
in Section \ref{focus} with the BPS framework.

\section{Selective Discipline-Specific Reviews of Bayesian
Forecasting\label{reviews}}

Having established the necessary details regarding the production of Bayesian
forecasts in general contexts, we now review how this general probabilistic
mechanism is employed to produce Bayesian\textbf{ }forecasts in several
important empirical fields. In order to produce a comprehensive and up-to-date
review of each area, a range of discipline experts have been invited to write
the various sections, with the authorship flagged in the section headings.
This means that the style of coverage differs somewhat across sections, as
suits the topic, and as fits with the perspective of the authors. However, we
have aimed to retain notation that (as far as possible) is both consistent
across sections and consistent with the notation used in the earlier parts of
the paper and in the technical appendix, plus to ensure that the basic layout
of all sections is the same. As noted earlier, other than in Section
\ref{mark} -- in which cross-sectional consumer choice data is modelled -- and
in Section \ref{elect}, in which spatial models are briefly referenced, time
series problems and forecasting are the primary focus.

\subsection{Macroeconomics\label{macro} (Florian Huber and Gary Koop)}

Central banks and other policy institutions routinely collect vast amounts of
time series data on key macroeconomic outcomes. One stylized fact is that
these data sets often display substantial co-movements and this calls for
modeling all these series jointly to produce accurate point and density
forecasts. This, however, leads to large-scale models that are prone to
overfitting, ultimately resulting in weak out-of-sample forecasting
performance. This helps explain the popularity of Bayesian methods for
macroeconomic forecasting. They can easily handle many parameters and, through
appropriate prior choice, deal effectively with questions related to model and
specification uncertainty in macroeconomic settings.

At a high level of generality, there are two modelling approaches used by
macroeconomic forecasters. The first uses reduced-form models and imposes
relatively little economic structure on the data. The second uses structural
models such as dynamic stochastic general equilibrium (DSGE) models that are
often estimated through Bayesian techniques; see, among many others,
\cite{ALV2007}, \cite{SW2007AER} and \cite{DS2016}.
However, reduced-form approaches have proved more popular and, in this
section, our focus will be on them.

As stated above, macroeconomists are typically interested in modeling the
joint evolution of a set of macroeconomic quantities. To set up a general
framework for understanding the types of models used for forecasting, assume
that an $M$-dimensional vector $\mathbf{y}_{t}$ is related to a $K$%
-dimensional vector of explanatory variables $\mathbf{x}_{t}$ through
\begin{equation}
\mathbf{y}_{t}=g(\mathbf{x}_{t})+\bm\varepsilon_{t},
\end{equation}
where $g:\mathbb{R}^{K}\rightarrow\mathbb{R}^{M}$ is a function and
$\bm\varepsilon_{t}$ is $\mathcal{N}(\bm0_{M},\bm\Sigma_{t})$.\footnote{Note
that the Gaussianity assumption is not essential; mixtures of Gaussian
distributions, for example, can be used to produce flexible error
distributions if deemed necessary (see, for example, \citealp{CHKMP2022}, and
\citealp{LP2022}).} This general specification nests most important
reduced-form models commonly used in macroeconomics and can be used to explain
the main issues that arise.

For instance, if $\mathbf{x}_{t}=(\mathbf{y}_{t-1}^{\prime},\dots
,\mathbf{y}_{t-p}^{\prime})^{\prime}$ contains $p$ lags of $\mathbf{y}_{t}$,
$g(\mathbf{x}_{t})=\mathbf{A}\mathbf{x}_{t}$ is a linear function with
$M\times K(=Mp)$ coefficient matrix $\mathbf{A}$, and $\bm\Sigma_{t}%
=\bm\Sigma$ is constant over time we have a standard vector autoregressive
(VAR) model. If we set $\mathbf{x}_{t}=\bm f_{t}$ with $\bm f_{t}$ denoting a
set of $Q\ll M$ latent factors and $g(\bm f_{t})=\bm\Lambda\bm f_{t}$ is
linear with $\bm\Lambda$ being an $M\times Q$ matrix of factor loadings and
$\bm f_{t}$ evolves according to some stochastic process (such as a VAR), we
end up with a dynamic factor model \citep[DFM, see][]{stock2011dynamic}.
Factor augmented VARs \citep{bernanke2005measuring} combine a VAR with a DFM.
The dependent variables in the VAR part of the model are a subset of
$\mathbf{y}_{t}$ plus a small number of factors.

Traditionally, VARs and factor models have been linear and homoskedastic. But
there is a great deal of empirical evidence in most macroeconomic data sets of
parameter change, both in the conditional mean and the conditional variance.
This can be accommodated through particular choices for $g$ and $\bm\Sigma
_{t}$. For the latter, stochastic volatility processes have proved
particularly popular. For the former, various parametric forms for $g$ lead to
time-varying parameter VARs (TVP-VARs) which assume that the coefficients of
the VAR evolve according to a random walk. But it is also worth noting that
there is an increasing literature which assumes $g$ is unknown and uses
Bayesian nonparametric methods to uncover its form (see, for example,
\citealp{kalli2018bayesian}, \citealp{adrian2021multimodality}, and \citealp{HUBER2020}).

If we set $M=1$ we obtain single-equation time series regressions, which are
particularly popular in inflation forecasting (e.g. based on the Phillips
curve). If we additionally set $x_{t}=1$ and allow for time-varying
parameters, we can obtain models such as the unobserved components stochastic
volatility (UCSV) model of \cite{SW2007} that is commonly used to forecast
inflation (for recent applications, see \citealp{chan2013new},
\citealp{stock2016core}, and \citealp{huber2021dynamic}).

This general framework defines a class of likelihood functions. As per the
outline in Section \ref{tute}, Bayesian forecasting involves multiplying a
chosen likelihood function by an appropriate prior to produce a posterior
which can be used to produce the predictive density. The choice of prior and
computational method used for posterior and predictive inference will be case
specific and we will have more to say about some interesting cases below. But
a few general comments are worth noting here. First, the choice of prior
matters much more in models such as the large VAR, which have a large number
of parameters relative to the number observations, than in models with fewer
parameters such as the UCSV model or the DFM. Second, for linear homoskedastic
models with conjugate priors analytical formulae for the posterior and the
one-step-ahead predictive density are available. For all other cases, MCMC
methods are available. These take the general form outlined in Section
\ref{sol}. However, as noted in Section \ref{intrac}, MCMC methods typically
do not scale well and can be computationally slow in models involving large
numbers of parameters (such as large VARs) or large numbers of latent states
(such as TVP-VARs). Thus, the focus of many recent papers has been on
developing either improved MCMC algorithms or approximate VB methods for
speeding up computation. Thirdly, our discussion so far focuses on forecasting
with a single model. In practice, it is common to find that forecasts improve
if many models are combined. Thus, either BMA or, alternatively, the methods
outlined in Section \ref{comb} are commonly used by macroeconomic forecasters.

With this general framework established, it is worthwhile to offer some
additional detail about some of the most important 21st century developments
and a discussion of how they have led to improvements in macroeconomic forecasting.

\paragraph{Large VARs}

Going back to early work such as \cite{DLS1984}, Bayesian VARs have been used
successfully in a variety of macroeconomic forecasting applications. Recently,
they have enjoyed even greater popularity due to the rise of the large VAR.
The pioneering large VAR paper was \cite{BGR2010}. Subsequently, dozens of
papers have used large VARs for macroeconomic forecasting (see, among many
others, \citealp{carriero2009forecasting}, \citealp{koop2013},
\citealp{carriero2015bayesian}, \citealp{GLP2015}, and
\citealp{hauzenberger2021combining}). Large VARs, involving dozens or even
hundreds of dependent variables, have been found to forecast well and improve
upon single-equation techniques and DFMs. Large VARs are heavily
over-parameterized and, thus, Bayesian prior shrinkage has been essential in
ensuring their forecasting success. We will discuss priors shortly, but at
this point we highlight the fact that the use of large Bayesian VARs has been
one of the major recent developments in macroeconomic forecasting.

\paragraph{Prior shrinkage in VARs}

Many different priors have been used with VARs. Traditionally, natural
conjugate priors in the Minnesota tradition were used since these allowed for
analytical posteriors and one-step-ahead predictives. Definitions of these
priors and discussions of their properties are available in standard sources
such as \cite{KK2010} and \cite{BEAR}. These priors are subjective and require
the user to select prior hyperparameters, most importantly those relating to
the strength of prior shrinkage. In recent years, a range of alternative
priors have been proposed which are more automatic, requiring fewer subjective
prior choices by the researcher. For instance, \cite{GLP2015} develop methods
for estimating shrinkage parameters in conjugate priors, thus avoiding the
need for their subjective elicitation. \cite{Chanasy2022} also uses a
conjugate prior and develops methods for selecting shrinkage parameters using
a prior which relaxes some of the restrictive assumptions of the Minnesota
prior. There are also a range of methods which automatically decide on the
optimal degree of shrinkage for each VAR coefficient. These are the
global-local shrinkage priors which are widely used with regressions and in
machine learning applications, and increasingly used with VARs.\footnote{They
are also used with DFMs to select the number of factors.} Global-local
shrinkage priors have the form
\[
a_{j}\sim\mathcal{N}(0,\psi_{j}\lambda),\quad\psi_{j}\sim f_{1},\quad
\lambda\sim f_{2},
\]
where $a_{j}$ is the $j^{th}$ VAR coefficient, $\lambda$ controls global
shrinkage since it is common to all coefficients, and $\psi_{j}$ controls
local shrinkage since it is specific to the $j^{th}$ coefficient. The
densities $f_{1}$ and $f_{2}$ are mixing densities and a large range of
choices of them have been proposed. One choice leads to stochastic search
variable selection, used with VARs in \cite{GSN2008}, \cite{koop2013} and
\cite{korobilis2013}, and many other references. Other choices lead to the
Dirichlet-Laplace prior used with VARs by \cite{KastnerHuber2021}, or the
{normal-gamma} and {horseshoe} priors used in \cite{huber2019adaptive} and
\cite{CROSS2020}; and there are many others. Since these priors are Gaussian
at the first layer of the hierarchy, textbook MCMC algorithms for all the VAR
parameters can be easily implemented.\footnote{In large VARs with global-local
shrinkage priors, MCMC methods can nevertheless be very slow, with much faster
VB methods developed in \cite{GEFANG2022}.}


\paragraph{Adding stochastic volatility (SV)}

The other main development that has had a tremendous impact on applied
macroeconomic forecasting in the 21st century is the development of models
such as VARs that incorporate parameter change and nonlinearity. Put simply,
the macroeconomic world is rarely linear and homoskedastic, and models that
relax these assumptions have been found to improve macroeconomic forecasting.
These improvements lie not only in point forecasts, but more importantly in
density forecasts. Given the increasing interest, by central banks and
academics alike, in issues such as forecast uncertainty and tail risk, the
fact that these new models produce more accurate predictive densities
increases their value.

A popular specification for VARs {with SV involves} factorizing the error
variance-covariance matrix as $\bm\Sigma_{t}=\mathbf{A}_{0}\mathbf{H}%
_{t}\mathbf{A}_{0}^{\prime}$ with $\mathbf{A}_{0}$ being a lower triangular
matrix with unit diagonals\footnote{$\mathbf{A}_{0}$ can also be time
varying.} and $\mathbf{H}_{t}=\text{diag}(e^{h_{1t}},\dots,e^{h_{Mt}})$ being
a diagonal matrix with log-volatilities evolving according to simple
stochastic processes such as independent random walks or AR(1) processes. In
an important contribution, \cite{clark2011real} considers a VAR-SV and finds
it to produce accurate point and density forecasts relative to homoskedastic
models, with gains being particularly pronounced using forecast metrics
involving the entire predictive density. Building on this insight, several
other researchers have analyzed the role of heteroskedasticity in
macroeconomic forecasting in VARs (see, for example,
\citealp{clark2015macroeconomic}, and \citealp{chiu2017forecasting}) and
confirm the result that using SV pays off when the focus is on obtaining
accurate density forecasts. However, a problem with the standard SV
specification is that the computational burden relative to homoskedastic VARs
is increased enormously. This makes it difficult to do Bayesian forecasting
with large VARs with SV. As a remedy, \cite{carriero2016common} propose a
simple common stochastic volatility (CSV) specification that assumes the shock
variances to be driven by a single common volatility factor, maintaining
conjugacy and thus leading to computationally efficient MCMC algorithms. They
acknowledge that this model is simplistic but show that it yields much more
accurate forecasts than homoskedastic VARs in a standard US macroeconomic
forecasting application.

To gain more flexibility, researchers have developed algorithms that allow for
estimating large VARs with $M$ independent SV processes.
\cite{carriero2019large} propose techniques that permit equation-by-equation
estimation of such VARs and thus render estimation of larger models with SV
feasible.
Modified versions of this algorithm form the basis of several recent papers
that combine large data sets with SV for macroeconomic forecasting (see, among
others, \citealp{huber2019adaptive}, \citealp{chan2021minnesota}, and \citealp{chan2021large}).

\paragraph{Adding time variation in the VAR coefficients}

The previous discussion has emphasized that capturing changing error variances
is key for obtaining precise forecasts. However, it may also be important to
allow for structural change in the VAR coefficients themselves. One popular
multivariate model that captures both changes in the VAR coefficients and
error variances is the TVP-VAR-SV model proposed in \cite{primiceri2005time},
which assumes that the VAR coefficients $\bm\beta_{t}=\text{vec}(\bm A_{t})$
are time-varying and evolve according to a multivariate random walk while
$\bm\Sigma_{t}$ is a multivariate SV process. This model is a multivariate
state space model which can be estimated using adaptations of the techniques
outlined in Section \ref{aug}. The innovations to the states govern the amount
of time variation in the parameters. Various shrinkage priors (often based on
the global-local shrinkage priors discussed above) have been proposed that
allow for a data-based decision as to whether time variation in a
corresponding coefficient is necessary or not. These priors are typically
elicited on the non-centered parameterization of the state space model
\citep[see][]{fruhwirth2010stochastic} and can help minimize overfitting
concerns and produce improved forecasts.

\cite{d2013macroeconomic} is an important early contribution to the
macroeconomic forecasting literature using TVP models. This paper uses a small
TVP-VAR with SV and shows that it produces more accurate point predictions,
outperforming simpler univariate benchmarks and constant parameter VARs. One
key shortcoming of this model, however, is that it only uses a small
information set. This has led to several researchers proposing new methods
that can be used in higher dimensions. Various approaches are possible,
including models that restrict the TVP process (e.g. by imposing a factor
structure, which allows for time variation in a large number of parameters to
be driven by a low number of factors, see \citealp{chan2020reducing}). As
mentioned above, shrinkage priors are used to keep the curse of dimensionality
in check. These priors are typically used after transforming the model to
allow for equation-by-equation estimation. Such approaches mean fairly
high-dimensional TVP-VARs can be estimated without risk of over-fitting, and
in a reasonable amount of time. MCMC-based forecasting with large TVP-VARs and
regressions is also an active field of research and different shrinkage
methods and advances in computation have led to improvements in the
forecasting performance of TVP models (see, among many others,
\citealp{HHKO2021}, and \citealp{HKO2021}). However, it is worth noting that
if computation does become a concern, approximate methods (e.g. using the VB
methods outlined in Section \ref{approx}) can be used. Approaches which avoid
the need for MCMC are developed in \cite{koop2013large} and
\cite{koop2018variational}. In the former paper the authors propose large
approximate TVP-VARs based on forgetting factors, whereas in the latter they
use VB techniques to forecast inflation with large TVP regression models.

\paragraph{Bayesian nonparametric VARs}

Up to this point we have assumed that the conditional mean function $g$ takes
a known form. However, it could be that the functional form is unknown.
Bayesian nonparametric techniques, such as Bayesian additive regression trees
\citep[BART, see][]{chipman2010bart}, Gaussian processes and kernel
regressions \citep{adrian2021multimodality} or infinite mixtures
\citep{kalli2018bayesian}, allow the researcher to uncover such unknown
functional forms and produce precise macroeconomic forecasts. In general, they
have had great success, but they have been found to be particularly useful in
studies that focus on the tails of predictive distributions or on the handling
of outliers such as the ones experienced during the pandemic (see, for
example, \citealp{HUBER2020}, and \citealp{clark2022forecasting}).


\cite{kalli2018bayesian} propose a nonparametric VAR that builds on an
infinite mixture model with the mixture weights being driven by the lagged
endogenous variables. They show, using US and UK data, that their model yields
competitive forecasts, with accuracy gains in terms of point and density
predictions increasing sharply for higher forecast horizons. \cite{CHKMP2022}
use BART-based VARs to perform tail forecasting of US output, unemployment and
inflation in {real time}, finding that nonparametric techniques work well in
the tails and for {higher-order} forecasts. With a particular focus on
predictive accuracy during the pandemic, \cite{HUBER2020} develop mixed
frequency nonparametric VARs and show that these models yield substantially
more precise nowcasts during the Covid-19 period.


\paragraph{Conclusions and further directions}

We have outlined how Bayesian methods have been used successfully for
macroeconomic forecasting. Most of the discussion has related to VARs, which
are a class of models where Bayesian methods have proved particularly popular.
But it is worth noting that empirically-relevant extensions (e.g. SV or TVP)
can be added to other multivariate time series models such as DFMs or FAVARs,
as can the VAR prior shrinkage methods (e.g. global-local shrinkage methods)
we have discussed. It is also worth noting that we have focused on models that
do not restrict the coefficients. However, restricted VARs are often used for
forecasting. For instance, vector error correction models (which impose
cointegrating restrictions) or multi-country VARs such as global VARs are
restricted VARs.

We have also focused on forecasting as opposed to the closely related field of
nowcasting. Mixed frequency VARs, which jointly model quickly-released,
high-frequency variables (e.g. monthly variables such as surveys, employment
and inflation) and slowly-released, low-frequency variables (e.g. quarterly
variables such as GDP), have proved very popular with nowcasters. Bayesian
methods are typically used with such models (see, for example,
\citealp{SS2015}, \citealp{HUBER2020}, \citealp{KMMP}, and
\citealp{stackedVAR}) and, in real-time nowcasting exercises they tend to
perform well.

\subsection{Finance\label{fin} (John Maheu, Worapree Maneesoonthorn and Gael
Martin)}

A pertinent question in financial analysis is whether the risks associated
with financial assets -- and the prices of those risks -- are predictable in
ways that are useful in applications such as portfolio allocation, risk
management and derivative pricing. With risk factors typically being
represented as latent distributional features of observable financial
variables, it follows that two key goals in the statistical analysis of
financial problems are: \textit{i)} The accurate prediction of latent
distributional features; and \textit{ii)} The development of complex,
non-linear state space models to underpin this prediction.

Both of these\textbf{ }goals lend themselves naturally to a Bayesian treatment
given, in turn, the automatic production of predictive \textit{distributions}
via the Bayesian paradigm, and the swathe of computational methods available
to estimate complex models -- most notably those with a latent variable
structure. In particular, the growth in financial derivatives markets from the
1990s onwards has generated the need to model the underlying asset as a
continuous time process, almost always augmented with a continuous time
process for the asset volatility, and often via a jump diffusion. Such models
-- whilst `convenient' in the sense of allowing for closed-form solutions for
derivative prices -- are challenging from a statistical point of view, given
that they typically need to be treated as a (discretized) non-linear
state-space model, and may require multiple sources of data to enable separate
identification of model parameters and risk premia. Estimation of and
forecasting with such models \textit{is }nevertheless computationally feasible
via Bayesian methods, with MCMC algorithms of one form or another forming the
backbone of the early treatments (\citealp{eraker2001}; \citealp{eraker2003};
\citealp{eraker2004}; \citealp{forbes2007}; \citealp{Johannes2009}).

We refer the reader to \cite{jacquier2011bayesian} and \cite{JOHANNES2010} for
comprehensive reviews of the application of Bayesian methods in finance up to
the first decade of the 21st century. The coverage includes, in short,
Bayesian approaches to: portfolio allocation, return predictability, asset
pricing, volatility, covariance, `beta' and `value at risk' prediction,
continuous time models (and discretized versions thereof), interest rate
modelling, and derivative (e.g. option) pricing. Our goal in the current
review is to outline the more recent advances that have evolved over the last
decade, in particular those that have exploited (in one way or another) new
methodological advances, new sources of data, and modern computational
techniques. In order, we shall briefly review: the use of diverse data sets,
including derivative prices and high-frequency measures of financial
quantities; the treatment of DGPs that are unavailable in closed form; the
analysis of high-dimensional models; and the application of non-parametric modelling.

\paragraph{Multiple sources of financial data}

It is now a well-established fact that the constant volatility feature of a
geometric diffusion process for a financial asset price is inconsistent with
both the observed dynamics in return volatility and the excess kurtosis and
skewness that characterizes the typical empirical return distribution; see
\cite{bollerslev:chou:kroner:1992} for an early review. The option pricing
literature supports this finding, with certain empirical regularities, such as
`implied volatility smiles', seen as evidence that asset prices deviate from
the geometric Brownian motion assumption that underlies the \cite{black1973}
option price \citep{bakshi1997,hafner2001,lim2005}. Hence, the 21st century
has seen the proliferation of many alternative specifications for asset
prices, and associated theoretical derivative prices, most of which are nested
in a general framework of (discretized) bivariate jump diffusion models for
the asset itself and its volatility. Allied with these developments has been
the growth in access to transaction-level `high-frequency' data -- in both the
spot and options markets -- which, in itself, has spawned new approaches to
inference and forecasting in the financial sphere.

The Bayesian literature has brought to bear on this problem the power of
computational methods -- both established, and more recent -- to enable the
multivariate state space models that have emerged from this literature to be
estimated, and probabilistic predictions of all dynamic variables -- the
return itself, volatility, random jumps (in either the return or the
volatility, or both), and various risk premia -- to be produced. With
reference to the generic notation for a state space model in (\ref{meas}) and
(\ref{state}), Bayesian approaches over the last decade can be categorized
according to the specification adopted for the (multivariate) measurement at
time $t$, $\mathbf{y}_{t}$ and, hence, for the (multivariate) state,
$\mathbf{z}_{t}$, being modelled and forecast. Some work exploits data from
both the spot and options market to predict volatility and its risk premia
(\citealp{maneesoonthorn2012}), and option prices\textbf{ }(\citealp{Yu2011};
\citealp{carverhill2022}\footnote{We note that whilst a time series model is
constructed in the case of these two references, the (out-of-sample)
prediction of option prices is across the cross section of strike prices and
maturities. We also make note of \cite{FULOP2019} who exploit spot and options
data to produce filtered estimates (as opposed to strictly out-of-sample
predictions) of latent volatility and price jump intensity.}); other work
combines `low-frequency' daily observations on returns with high-frequency
measures of volatility and/or price jumps to predict (in some combination)
returns, volatility, and the size and occurrence of price jumps
\citep{jin2013modeling,maneesoonthorn2017,frazier2019approximate}; whilst
further work combines daily returns with futures prices in predicting various
financial quantities of interest \citep{FILECCIA2018,gonzato2021}.

\paragraph{Financial models that are `unavailable'}

All but one of the papers cited in the previous paragraphs share a common
feature - namely, a DGP that can be expressed as a probability density (or
mass) function. With reference to (\ref{aug_post}), it is the availability of
a closed form for $p(\mathbf{y}_{1:T},\mathbf{z}_{1:T}|\boldsymbol{\theta
})=p(\mathbf{y}_{1:T}|\mathbf{z}_{1:T},\boldsymbol{\theta})p(\mathbf{z}%
_{1:T}|\boldsymbol{\theta})$, that renders feasible the MCMC methods used in
the said works. In contrast, \cite{frazier2019approximate} adopt a process for
the latent log-volatility that is driven by an $\alpha$-stable innovation,
such that $p(\mathbf{z}_{1:T}|\boldsymbol{\theta})$ is unavailable, and MCMC
infeasible as a consequence. Instead, ABC is adopted for inference, and an
approximate predictive of the form of (\ref{eq12}) produced instead. In
addition to providing theoretical validation of the approach, the authors
demonstrate, in range of different simulation settings, that despite
inaccuracy at the posterior level, the approximate predictive is always a very
close match to the exact predictive. Related work in which an ABC method is
used to conduct forecasting appears in \cite{canale2016},
\cite{konkamking2019}, \cite{VIRBICKAITE2020} and \cite{pesonen2022}. ABC
treatment of a conditional likelihood for a time series of financial returns,
$p(\mathbf{y}_{1:T}|\mathbf{z}_{1:T},\boldsymbol{\theta})$, that is
unavailable in closed form is also investigated in \cite{CREEL2015},
\cite{martin2019auxiliary} and \cite{chakraborty2022modularized}, with
\cite{chakraborty2022modularized} proposing a modularized version of ABC. For
other recent Bayesian treatments of intractable models of this sort that
continue to exploit MCMC principles (with or without an ABC component), see
\cite{vankov2019filtering} and \cite{muller2021estimation}.\footnote{The
citation of \cite{CREEL2015}, \cite{martin2019auxiliary},
\cite{vankov2019filtering} and \cite{muller2021estimation} is relevant to this
review, despite these references not having an \textit{explicit} component on
forecasting.}

\paragraph{Large financial models}

Thus far, we have reviewed\textbf{ }Bayesian treatments of models for single
financial assets. That is, the\textbf{ }models may have specified multiple
latent components, and potentially multiple measurements, but they\textbf{
}still aim to explain (and forecast) quantities related to a \textit{single}
asset. Models for multiple assets\textbf{ }are also critically important in
financial applications, with the relationship between financial assets
determining the extent to which diversification can be achieved, as well as
how risks permeate the various sectors of the financial market. Indeed,
Bayesian methods are particularly suitable for dealing with such multivariate
models, since the dimensionality of $\mathbf{z}_{1:T}$ is typically much
larger than that of $\mathbf{y}_{1:T}$ and, hence, challenging to deal with
via any other means.

\cite{chib2009Review} provide an early review of the Bayesian analysis of
multivariate SV models, with all work up to this point utilizing traditional
MCMC techniques, and the statistical and predictive analysis limited to
relatively low-dimensional systems (up to ten assets). Subsequent work has
focused on the development of more flexible multivariate distributions
\citep{nakajima2017bayesian}, and the use of sparse factor structures and
shrinkage priors in constructing larger-dimensional models
\citep{zhou2014bayesian,kastner2017efficient,BASTURK2019}. More
recently, with the advances made in VB methods, inference and prediction in
very large-dimensional financial models is now possible
(\citealp{gunawan2021variational}; \citealp{Chansv2022};
\citealp{frazierssm2021}; \citealp{quiroz2018gaussian}; \citealp{zhangw2023}). There is also a
growing interest in the prediction of co-movements of various sorts, with:
\cite{bernardi2015bayesian} predicting the interdependence between U.S. stocks
with Bayesian time-varying quantile regressions; \cite{geraci2018measuring}
capturing and predicting the interconnectedness of financial institutions
through Bayesian time-varying VARs; and \cite{alexopoulos2022bayesian}
modelling and predicting common jump factors in a large panel of financial returns.

\paragraph{Bayesian nonparametric modelling in finance \label{BNP}}

As noted, simple parametric assumptions such as additive Gaussian innovations
are inconsistent with the stylized features of financial data. Whilst more
suitable non-Gaussian/non-linear models can be built (as highlighted above),
Bayesian nonparametric modelling allows for further flexibility via the
incorporation of Dirichlet process mixture (DPM) structures. Such an approach
has been shown to provide robustness to distributional assumptions and can
improve point forecasts, but the main gain has been significant improvements
in the accuracy of predictive densities, and of risk measures derived from
those densities. The advancement of the literature in this direction has been
aided by the stick-breaking representation \citep{sethuraman1994constructive}
and the introduction of the slice sampler \citep{walker2007sampling,kalli2011slice}.

\cite{jensen2010bayesian} introduce an extension to a standard SV model to
capture the unknown return innovation distribution via a DPM. The DPM
specification has also\textbf{ }been inserted into other popular models in
finance, with: \cite{jensen2014estimating} adopting a DPM to jointly model the
return and future log-volatility distribution; \cite{delatola2013bayesian}
capturing the so-called leverage effect; \cite{AUSIN2014350} applying
a\textbf{ }DPM to univariate GARCH models; and \cite{Kalli-Griffin:2015} using
Bayesian nonparametric modelling to aggregate autoregressive processes to
produce an SV model with long-range dependence. Extensions to multivariate
financial models have also occurred: in a multivariate GARCH setting
in\textbf{ }\cite{Jensen-Maheu:2013}; and in a Cholesky-type multivariate SV
model in \cite{ZTW:2020}.

A potential drawback of the DPM model is that it neglects time dependence in
the unknown distribution. An important extension of the DPM prior is the
hierarchical Dirichlet process of \cite{teh2006hierarchical}, which allows for
the construction of a prior for an infinite hidden Markov model (IHMM), which
allows for time dependence in a flexible manner. The introduction of
the\textbf{ }beam sampler of \cite{van2008beam}, which extends the slice
sampler, renders conventional posterior sampling methods for finite-state
Markov switching models \citep{CHIB1996} feasible in the IHMM. The IHMM
structure has been used to model the univariate GARCH distribution
\citep{dufays2016infinite}, and the multivariate GARCH distribution
\citep{Robin:2022}; and to provide a nonparametric model for realized
measures, including realized covariance matrices
\citep{jin2016bayesian,liu-maheu:2018,jin2019bayesian}, with all papers
documenting very large improvements in density forecast accuracy from the
IHMM. Other applications of the IHMM include: \cite{shi-yong:2016}, who use
the IHMM to date and forecast speculative bubbles, and who also adopt a
version with GARCH effects; \cite{yang2019}, who studies the relationship
between stock returns and real growth with a multivariate IHMM model; and,
more recently, \cite{JIN2022302}, who employ the DPM prior in the infinite
Markov pooling of predictive distributions, with forecasting applications to
interest rates, realized covariances and asset returns. Other approaches to
time dependence in Bayesian nonparametrics for finance include
\cite{griffin2011stick}, who introduce a time-dependent stick breaking process
in a general setting and develop an SV model for returns. More recently,
\cite{SKL:2020} use a weighted DPM to forecast return distributions, while
\cite{SHAMSI:2021} allows for lagged covariates to impact the weights in the
DPM model through a probit stick-breaking process.

\subsection{Marketing (Rub\'{e}n Loaiza Maya and Didier Nibbering)\label{mark}%
}

Bayesian methods are applied to a wide range of marketing problems; see
\cite{rossi2003bayesian} for a review of the early literature. More recently,
these methods have been increasingly used for the purpose of prediction, for
instance in customer choice behaviour
\citep{toubia2019extracting,araya2022identifying}, customer demand
\citep{posch2022bayesian}, customer satisfaction \citep{mittal2021improving},
dynamic pricing \citep{bastani2022meta}, advertising effectiveness
\citep{danaher2020advertising,loaiza2022fast} and recommender systems
\citep{ansari2018probabilistic}. Given the large variety of marketing
applications, we focus in this section on the modelling of customer choice to
illustrate the key principles of Bayesian prediction in marketing problems.

A common problem in marketing is that of setting the price level
of a set of products so that total profits are maximized. To estimate these
optimal prices, predictions of how customers will react to price changes are
crucial. Predictions of customer choices under different marketing
environments can be constructed by choice models. These models are estimated
using data about the product choices of customers in the marketplace, a
survey, an experiment, etc. \citep{rossi2012bayesian}.

An example of a prediction of interest in this context is the predicted
purchase probability of a customer for a particular product as a function of
its own price or the price of another product. The predicted purchase
probability can be constructed for a customer for which only a few choices are
observed, or for a new customer for which we do not observe choices in the
data.\footnote{Although this section, as noted in the Introduction, focuses on
prediction using cross-sectional data, choice models can also be applied to
the forecasting of future choice probabilities by using time series data
\citep{mccormick2012dynamic} or panel data
\citep{gilbride2004choice,terui2011effect}.}



The two most popular models used to predict choice behaviour are the
multinomial logit and multinomial probit models. The multinomial logit model
imposes the independence of irrelevant alternatives (IIA) property
\citep{mcfadden1989method}, which means that it cannot capture general
substitution patterns among choice alternatives. The IIA property of this
model can be relaxed under certain assumptions by extending the multinomial
logit model to a nested logit model
\citep{poirier1996bayesian,lahiri2002bayesian} or\ a random parameter logit
model \citep{train2009discrete}.

On the other hand, the multinomial probit model does not impose the IIA
property, and as such is commonly used in the analysis of economic choice
behaviour, where complementary and substitution effects are important. For
instance, the multinomial probit model has been recently used in the analysis
of car choices \citep{karmakar2021understanding}, grocery brand choices
\citep{miyazaki2021dynamic}, employment choices \citep{mishkin2021gender}, and
car parking choices \citep{paleti2018generalized}.
The remainder of this section presents a review of Bayesian prediction based
on the multinomial probit model.

\paragraph{Multinomial probit model specification}

The variable of interest is $y_{i}\in\{0,1,2,\dots,J\}$, which indicates the
choice made by individual $i$ among a set of $J+1$ alternatives. This choice
is modeled to be conditional on a set of $J$ latent utilities $\mathbf{z}%
_{i}=\left(  z_{i1},\dots,z_{iJ}\right)  ^{\prime}$, so that the conditional
pmf is defined as
\begin{equation}
p(y_{i}|\mathbf{z}_{i})=%
\begin{cases}
I\left[  z_{iy_{i}}=\text{max}(\mathbf{z}_{i})\right]   & \text{ if
}\text{max}(\mathbf{z}_{i})>0,\\
I[y_{i}=0] & \text{ if }\text{max}(\mathbf{z}_{i})\leq0,
\end{cases}
\label{eq:y_ik}%
\end{equation}
where $p(y_{i}|\mathbf{z}_{i})=\text{Pr}(Y_{i}=y_{i}|\mathbf{z}_{i})$,
$z_{iy_{i}}$ is the $y_{i}$-th element of $\mathbf{z}_{i}$, with $y_{i}>0$,
and $I[A]$ is one if statement $A$ is true and zero otherwise. The base
category $j=0$ is one of the choice alternatives, which is selected \textit{a
priori}. The base category is observed whenever all the latent utilities are
less than zero.

The utilities are expressed in terms of $r$ predictors via a linear Gaussian
model,
\begin{equation}
p(\mathbf{z}_{i}|X_{i},\boldsymbol{\theta})=\phi_{J}\left(  \mathbf{z}%
_{i};X_{i}{\boldsymbol{\beta}},\Sigma\right)  , \label{eq:x}%
\end{equation}
where $\phi_{J}\left(  \mathbf{z};\boldsymbol{\mu},C\right)  $ denotes a
$J$-variate normal density with mean $\boldsymbol{\mu}$ and covariance matrix
$C$, $X_{i}$ a $J\times r$ matrix of predictor values, $\boldsymbol{\beta}$ an
$r$-dimensional vector of coefficients, and $\Sigma$ a covariance matrix that
captures complementary and substitution effects between the choice alternatives.

Combined, \eqref{eq:y_ik} and \eqref{eq:x} give rise to the augmented
likelihood function of the multinomial probit model
\begin{equation}
p(\mathbf{y},\mathbf{z}|\boldsymbol{\theta},X)=\prod_{i=1}^{n}p(y_{i}%
|\mathbf{z}_{i})p(\mathbf{z}_{i}|{X}_{i},\boldsymbol{\theta}),
\end{equation}
where $\boldsymbol{\theta}=\{\boldsymbol{\beta},\Sigma\}$, $\mathbf{y}%
=\{y_{i}\}_{i=1}^{n}$, $\mathbf{z}=\{\mathbf{z}_{i}\}_{i=1}^{n}$, and
$X=\{X_{i}\}_{i=1}^{n}$, with $n$ the total number of individuals.
%
For a given prior distribution $p(\boldsymbol{\theta})$, the augmented
posterior distribution of the model is given as
\begin{equation}
p(\boldsymbol{\theta},\mathbf{z}|\mathbf{y},X)\propto p(\mathbf{y}%
,\mathbf{z}|\boldsymbol{\theta},X)p(\boldsymbol{\theta}). \label{Eq:post}%
\end{equation}
\cite{albert1993bayesian} were the first to propose the use of data
augmentation (see Section \ref{aug} herein) for conducting Bayesian analysis
of the multinomial probit model.

\paragraph{The predictive distribution}

Consider now an individual $s$, with predictor values $X_{s}$, whose choice
behaviour we would like to predict. The predictive for individual $s$, can be
written as
\begin{equation}
p({y}_{s}|X_{s},\mathbf{y},X)=\int_{\Theta}\int_{\mathbf{z}_{s}}p({y}%
_{s}|\mathbf{z}_{s})p(\mathbf{z}_{s}|\boldsymbol{\theta},X_{s})d\mathbf{z}%
_{s}\int_{\mathbf{z}}p(\boldsymbol{\theta},\boldsymbol{z}|\mathbf{y}%
,X)d\mathbf{z}\,d\boldsymbol{\theta}, \label{eq:predictive}%
\end{equation}
from which the predictive choice probabilities $\text{Pr}(Y_{s}=j|X_{s}%
,\mathbf{y},X)=p(j|X_{s},\mathbf{y},X)$ can be constructed. The specification
and computation of the predictive distribution in \eqref{eq:predictive} poses
three key challenges.

First, $p({y}_{s}|\mathbf{z}_{s})$ requires a choice of base category. This
choice affects the prior predictive choice probabilities, and hence the
(posterior) predictive choice probabilities can be sensitive to the choice of
base category; see \cite{burgette2012trace}. \cite{burgette2021symmetric}
propose a symmetric prior specification to address this problem. The
parameters $\boldsymbol{\theta}$ are not identified under this prior, but this
does not affect the predicted probabilities.
%

Second, the parameters $\boldsymbol{\theta}$ lack scale identification, as
$p(y_{i}|\mathbf{z}_{i})=p(y_{i}|c\mathbf{z}_{i})$ for any positive scalar
$c$. Different solutions have been proposed to fix the scale, all based on a
constraint on the specification of $\Sigma$. For instance,
\cite{mcculloch2000bayesian} fix the first leading element of $\Sigma$ to
unity. This approach is sensitive to the ordering of the choice categories in
the model. \cite{burgette2012trace} fix the trace of $\Sigma$, which is
invariant to the way in which the choice categories enter the model.
%

Third, the computation of $p({y}_{s}|X_{s},\mathbf{y},X)$ involves the
evaluation of the integrals over the latent utilities in $\mathbf{z}_{s}$ and
$\mathbf{z}$. Since no analytical solution for these integrals is available,
they are solved with MCMC sampling steps. The latent utility of each choice
category is sampled from a univariate truncated normal, conditional on the
latent utilities for all the other choice alternatives, for each individual
\citep{mcculloch1994exact}. Conditional on the draws for the latent utilities,
sampling $\boldsymbol{\beta}$ from its full conditional is straightforward.
Generating from the conditional distribution of $\Sigma$ is nonstandard as the
scale restrictions on $\Sigma$ have to be taken into account.

\paragraph{Scalable Bayesian prediction}

In addition to the challenges delineated above, it is difficult to scale
$p({y}_{s}|X_{s},\mathbf{y},X)$ to problems with large choice sets or a large
number of observations. Recent advances in the computation of the predictive
have focused on tackling the scalability issues in $J$ and $n$, as we discuss below.

When considering a full covariance matrix specification for $\Sigma$, the
total number of parameters increases quadratically with $J$. For problems with
large choice sets and small samples, this implies that the ratio of total
number of parameters to total number of observations is large, making it
difficult to construct accurate predictions. \cite{loaiza2021scalable} propose
a spherical transformation of the covariance matrix of the latent utilities
that imposes a parsimonious factor structure and a trace restriction. As a
result, the total number of parameters grows only linearly with $J$. The
authors demonstrate that this parsimonious structure leads to improved
predictive performance over full covariance matrix specifications.

Additionally, as noted above, the construction of the predictive entails
evaluation of the integral over the latent utilities $\boldsymbol{z}$.
Although MCMC is able to solve this integral, it does so by generating the
utility vector for each individual from a multivariate truncated normal, which
is a computationally costly exercise
\citep{mcculloch1994exact,botev2017normal}. This renders MCMC algorithms
impractical for problems where a large $n$ is considered.

VB can be employed to tackle problems with large $n$. Adapting the generic
descriptions of VB\ in Section \ref{approx} and Appendix \ref{A7}, the
application of VB in this setting considers the class of approximating
densities $\mathcal{Q}$ with elements $q_{{\lambda}}(\boldsymbol{\theta
},\mathbf{z})\in\mathcal{Q}$, indexed by a variational parameter vector
$\boldsymbol{\lambda}$. The exact augmented posterior is approximated by
$q_{\hat{\lambda}}(\boldsymbol{\theta},\mathbf{z})$ with an optimal
variational parameter vector equal to
\begin{equation}
\hat{\boldsymbol{\lambda}}=\argmin_{\boldsymbol{\lambda}\in\Lambda}%
\text{KL}\left[  q_{{\lambda}}(\boldsymbol{\theta},\mathbf{z}%
)|p(\boldsymbol{\theta},\mathbf{z}|\mathbf{y},X)\right]  ,
\end{equation}
where KL denotes the Kullback-Leibler divergence. The variational predictive
is then constructed as%
\[
\hat{p}_{\lambda}({y}_{s}|X_{s},\mathbf{y},X)=\int_{\Theta}\int_{\mathbf{z}%
_{s}}p({y}_{s}|\mathbf{z}_{s})p(\mathbf{z}_{s}|\boldsymbol{\theta}%
,X_{s})d\mathbf{z}_{s}\int_{\mathbf{z}}q_{\hat{\lambda}}(\boldsymbol{\theta
},\mathbf{z})d\mathbf{z}\,d\boldsymbol{\theta}.
\]
Calibration of the variational approximation requires a scale-identified
expression for $p(\mathbf{y},\mathbf{z}|\boldsymbol{\theta},X)$. To achieve
this, \citet{girolami2006variational} consider an identity matrix covariance
structure, while \citet{fasano2022class} fix $\Sigma$ at predetermined values.
\citet{loaiza2022mnpvb} propose a method for a multinomial probit model with a
factor covariance structure. This method uses the hybrid variational
approximation $q_{{\lambda}}(\boldsymbol{\theta},\mathbf{z})=q_{{\lambda}%
}(\boldsymbol{\theta})p(\mathbf{z}|\mathbf{y},\boldsymbol{\theta},X)$
introduced by \citet{loaiza2022fast}.

\subsection{Electricity Pricing and Demand (Anastasios
Panagiotelis)\label{elect}}

Forecasting in electricity markets is critical for efficient day-to-day
operation of power grids, long-term planning of infrastructure and
increasingly, at a disaggregated level, for the management of smart grids.
This section will cover forecasting electricity prices, electricity
load/demand and generation by source of power, primarily wind and solar.
Hereafter these problems will collectively be referred to as `electricity
forecasting'. Motivations for electricity forecasting can be found in general
reviews such as \cite{Wer2014} for price forecasting, \cite{LinEtal2019} for
load forecasting, \cite{Ant2016} for solar power forecasting and
\cite{GieKar2017} for wind power forecasting. These reviews indicate that the
majority of work in electricity forecasting does not employ a Bayesian
approach; however notwithstanding this, Bayesian methods have found success in
the field.

There are very few instances of Bayesian forecasting in electricity markets
that predate the early 2000s, although we now cover some notable exceptions.
\citet{Bun1980} consider the case of updating load forecasts in an online
fashion by computing a Bayesian model average of load profiles of a cloudy and
a sunny day. Meanwhile, Bayesian VARs have been used by \cite{Gun1987},
\cite{BecSol1994} and \cite{JouEtal1995} to forecast energy demand, nuclear
power generation and demand prices and consumption respectively. A Bayesian
VAR shrinks autoregressive coefficients to either a random walk or white noise
depending on whether data are stationary or non-stationary and was popularized
in macroeconomics by \cite{DLS1984} (see also Section \ref{macro}). The
performance of Bayesian VARs in early electricity forecasting applications is
mixed; \cite{BecSol1994} find evidence in favour of Bayesian autoregression,
\cite{JouEtal1995} find that Bayesian VARs are effective for forecasting
demand, but not price, while \cite{Gun1987} does not find any improvement at
all from using Bayesian VARs rather than conventional autoregressive
integrated moving average (ARIMA) models.

With the advent and popularization of MCMC methods, Bayesian forecasting has
begun to find greater success in the field of electricity forecasting. In the
literature of roughly the past two decades, there are three common major
motivations for using Bayesian forecasting, two of which have antecedents in
the earlier literature. The first is the use of `Bayesian models'\footnote{By
a `Bayesian model' we generally mean a model with a prior and likelihood
estimated by Bayesian inference. Bayesian methods for finding tuning
parameters such as the automatic relevance determination
\citep[see][for an example in electricity forecasting]{HipTay2010}, and
Bayesian optimisation lie beyond the scope of this section.}, which have now
grown well beyond Bayesian VARs to include models with latent volatilities,
models with a spatial dimension, and Bayesian neural networks. The second is
the use of BMA for forecast combination. The third is the production of full
probabilistic forecasts via Bayesian computation. These are now each discussed
in turn.

\paragraph{Bayesian models}

The structure inherent in many electricity forecasting problems provides a
motivation for the innovative use of priors to improve forecasting accuracy.
Although the early literature cited before found somewhat ambiguous results
when comparing Bayesian VARs to classical alternatives, more recent work finds
evidence in favour of a Bayesian approach; see \cite{RavEtal2015} for point
forecasts and \cite{GiaEtal2020} for both point and density forecasts. An
important aspect of this work is the exploitation of the intraday nature of
the data, since typically hourly prices are stacked in a VAR model. The
intraday structure lends itself to priors that shrink parameters corresponding
to consecutive hours of the day that are close to one another. An early
application of this approach can be seen in \cite{CotSmi2003}.

Since electricity data are increasingly available not only at a high temporal
frequency but also at a high spatial resolution, there are further examples in
the literature of using priors to exploit neighbourhood structure. Examples
include \cite{OhtEtal2010} who use spatial ARMA processes to predict
electricity load in nine Japanese regions, and \cite{GilEtal2019} who use
spatio-temporal Gaussian processes to forecast residential-level electricity
demand. Even where spatial information is unavailable, hierarchical models
estimated using Bayesian methods have been used to produce disaggregate energy
demand forecasts; examples can be found in \cite{MorNak2014} and
\cite{WanEtal2017} who use Gaussian processes, and \cite{GriEtal2021} who use
regression. Informative hierarchical priors have been used in instances where
data sets are small in size, or unavailable; for example, \cite{PezEtal2006}
elicit priors for future trajectories of temperature in the winter using past
observations, and \cite{LauEtAl2015} elicit priors for the electricity demand
of `non-metered' households using data on `metered' households.

While the aforementioned examples take a Bayesian approach to exploit the use
of priors in novel ways, another strain of the Bayesian forecasting literature
is based on estimating models with latent variables. Examples in electricity
forecasting include a latent jump process for price spikes \citep{ChaEtal2014}
and SV models (\citealp{Smi2010}; \citealp{KosKos2019}). Also, in recent
years, Bayesian analysis of machine learning models has become increasingly
popular. This includes neural network models
\citep{BruEtal2019,GhaEtal2019,CapEtal2020}, where VB is typically used. Also,
Bayesian regression trees (see Section \ref{macro}) have been applied to
electricity forecasting by \cite{NatEtal2011} and \cite{AliEtal2019}, who find
that they outperform non-Bayesian counterparts. Finally, there is an extensive
literature on using Bayesian networks for forecasting in energy; see
\cite{AdeEtal2020} for a review of these methods in forecasting wind generation.

\paragraph{Bayesian model averaging (BMA)}

As noted earlier, the importance of forecast combination is widely appreciated
in the forecasting literature. Whilst, as highlighted in Section \ref{comb},
many different Bayesian approaches to forecast combination have now been
explored, BMA remains a very important method in the sphere of electricity
forecasting. As described in Section \ref{tute}, BMA uses posterior model
probabilities as combination weights. Whenever the choice of model is
parameterized, the predictive density has an interpretation as a forecast
combination. Examples include \cite{Smi2000} who combines forecasts from
regression models that include different predictor sets, and \cite{PanSmi2008}
who average over models with different combinations of skew and symmetric
marginal distributions.

It is also common in the electricity forecasting literature to produce point
forecasts from different models and then combine these using BMA as a
post-processing step. This approach grew out of research combining ensembles
of forecasts from numerical weather predictions (NWPs)
\citep{RafEtal2005,SloEtal2010}. In the NWP setting, forecasts are the outputs
of deterministic physical models. Statistical models are then formed by
assuming that for $k=1,\dots,K$, $p(y_{t}|a_{k},b_{k},f_{k},\sigma
^{2},\mathcal{M}_{k})\sim N(a_{k}+b_{k}f_{k},\sigma^{2})$, where $f_{k}$ is
the $k^{th}$ NWP and $a_{k}$, $b_{k}$ and $\sigma^{2}$ are additional
parameters. These statistical models are then combined using the usual BMA
machinery described by (\ref{mod_av}), with the key distinction being that
posterior model probabilities are replaced with $p(M_{k}|y_{T-L+1:T})$, where
$L$ is the length of the window. Uncertainty over $a_{k}$, $b_{k}$ and
$\sigma^{2}$ is integrated out in the usual way, and there are no additional
parameters since the $f_{k}$ are obtained deterministically. This approach has
been used in energy forecasting by \cite{Coe2006}, who motivate forecasting
rainfall as a input into forecasting generation from hydroelectric dams, and
\cite{Du2018} who uses wind forecasts to predict generation from wind farms.

The work of \cite{RafEtal2005} has been subsequently extended to the case
where the forecasts $f_{k}$ are not the outputs of deterministic physical
models but are point forecasts from statistical models, each with their own
unknown parameters. For example \cite{NowEtal2014} adopt the approach of
\cite{RafEtal2005} but where the $f_{k}$ are obtained from statistical time
series models with parameters estimated using frequentist techniques. {This
approach is not fully Bayesian} (despite being referred to as BMA in the
literature),{ since although the model average integrates over the uncertainty
in $a_{k}$, $b_{k}$ and $\sigma^{2}$ it does not integrate over uncertainty in
the parameters of the underlying time series models used to generate the point
forecasts $f_{k}$.\footnote{The same point does not apply when combining
ensembles from NWPs since the forecasting models are deterministic.}} In a
similar vein, \cite{HasEtal2015} and \cite{RazEtal2017} combine electricity
load forecasts from different neural networks.

\paragraph{Probabilistic forecasting}

A common motivation for taking a Bayesian approach is the ease with which the
computational machinery of MCMC or approximate methods produces a full
predictive density rather than only point forecasts. Key operational decisions
in electricity forecasting depend on quantities other than the predicted mean;
see \cite{NowWer2018} and references therein for discussion. While the
importance of probabilistic forecasting is often highlighted in Bayesian
papers it is not always the case that forecasts are evaluated in a way that
assesses the quality of the full predictive distribution\footnote{We note that
in some cases this is challenging; for example for long-run forecasts as in
\cite{daSEtal2019}.}. For example, often probabilistic forecasts are
summarized by prediction intervals, and the empirical coverage of these
intervals used as a means of checking model quality; for an early example see
\cite{PezEtal2006}, and more recently \cite{WanEtal2017} and \cite{KosKos2019}%
, where the latter show that Bayesian methods compare favourably to
non-Bayesian alternatives for forecasting electricity prices.
\cite{KosKos2019} also evaluate $\alpha$-level quantile forecasts $\hat{q}%
_{t}$ using the pinball loss,
\[
L_{\alpha}(y_{t},\hat{q}_{t})=\alpha(y_{t}-\hat{q}_{t})I[y_{t}\geq\hat{q}%
_{t}]+(1-\alpha)(\hat{q}_{t}-y_{t})I[y_{t}<\hat{q}_{t}]\,.
\]
\cite{YanEtal2019} and \cite{SunEtal2019} also use pinball loss to evaluate
forecasts of residential-level load (net of solar PV generation in the latter case).

However, the use of scoring rules \citep{gneiting2007strictly} and, hence, the
explicit recognition of the distributional form of the forecasts, \textit{is}
becoming increasingly popular as a means of evaluating predictive
distributions in both Bayesian and non-Bayesian electricity forecasting. The
{continuously ranked probability score }(\citealp{gneiting2007strictly}) is
particularly amenable to Bayesian inference since it is usually approximated
using a Monte Carlo sample from the predictive density. For an early example
of its use in Bayesian electricity forecasting see \cite{PanSmi2008}; for
later examples, see \cite{BraDeF2015}, \cite{BruEtal2019} and
\cite{GiaEtal2020}. Other scoring rules are less commonly used in the Bayesian
electricity forecasting literature, although \cite{OhtEtal2010}, where\textbf{
}the log score is used, is a notable exception.

\section{In Summary\label{concl}}

Bayesian forecasting is underpinned by a single core principle: uncertainty
about the future value of a random variable is expressed using a probability
distribution, where the form of that distribution reflects -- in turn --
uncertainty about all other unknowns on which the investigator chooses not to condition.

While this principled approach to forecasting is arguably one of the most
compelling features of the paradigm, the challenge has, potentially, been in
the \textit{implementation} of Bayesian forecasting: namely, computing the
expectation that defines the predictive distribution, most particularly when
accessing (draws from) the posterior itself is difficult. And as models have
become larger and more challenging, and as data sets have grown `bigger', this
problem of accessing the exact posterior has only increased. However, as this
review has demonstrated, the expansion of the forecasting problems being
tackled has gone hand-in-hand with the development of new and improved
computational methods designed expressly to access challenging posteriors, and
in a reasonable computing time. Notably, when it comes to accurate
forecasting, somewhat crude \textit{approximations} of the posterior have been
found to still yield accurate predictions; meaning that Bayesian forecasting
remains viable for large and complex models for which approximate computation
of posteriors is the only feasible approach.

The more fundamental problem of model misspecification can also be managed, by
moving away from the conventional likelihood-based Bayesian updating and
allowing forecast accuracy itself -- and its link to the future decisions that
depend on that accuracy -- to drive the updating. This, in turn, ensures that
forecasts are `fit for purpose', \textit{despite} the inevitable
misspecification of the forecasting model. Allied with the computational power
that now drives the Bayesian engine, this ability to generalize the paradigm
beyond its traditional links with the likelihood principle is a potent, if not
yet fully realized, force in forecasting.

\bibliographystyle{apalikeit}
\bibliography{Bayes_comp_forecast_corrected}

\appendix

\section{Further Computational Details}

\label{app:algos}

\subsection{Gibbs sampling\label{A1}}

Under the required regularity conditions (see \citealp{tier:1994}) the Gibbs
sampler yields a Markov chain with invariant distribution,
$p(\boldsymbol{\theta}\mathbf{|y}_{1:T})$, via a transition kernel that is
defined as the product of full conditional posteriors\textit{ }associated with
the joint. For the case of $\boldsymbol{\theta}$ partitioned into $B$ mutually
exclusive blocks, $\boldsymbol{\theta}=(\boldsymbol{\theta}_{1}^{\prime
},\boldsymbol{\theta}_{2}^{\prime},...,\boldsymbol{\theta}_{b}^{\prime
},...,\boldsymbol{\theta}_{B}^{\prime})^{\prime}$, the steps of the Gibbs
algorithm are given in Algorithm \ref{Gibbsalg1}.

\begin{algorithm}
	\caption{Gibbs Sampling Algorithm}
	\label{Gibbsalg1}
	\begin{algorithmic}
		\STATE Specify an initial value $\bt^{(0)}$ and partition the parameter set into $B$ mutually exclusive blocks.
		\FOR{$i=1,\dots,M$}	
		\FOR{$b=1,\dots,B$}
		\STATE Draw $\bt_b^{(i)}\sim p_b(\bt_b|\bt_1^{(i)},\dots,\bt_{b-1}^{(i)},\bt_{b+1}^{(i-1)},\dots,\bt_{B}^{(i-1)},\mathbf{|y}_{1:T})$
		\ENDFOR
		\ENDFOR
		\STATE Return a sample of draws from $p(\bt\mathbf{|y}_{1:T})$.
	\end{algorithmic}	
\end{algorithm}

\subsection{MH-within-Gibbs sampling\label{A2}}

In Algorithm \ref{MHwithinGibbsalg1} we provide the generic steps of the
so-called `MH-within-Gibbs' algorithm, for the case of $\boldsymbol{\theta}$
partitioned into $B$ mutually exclusive blocks, $\boldsymbol{\theta
}=(\boldsymbol{\theta}_{1},\boldsymbol{\theta}_{2},...,\boldsymbol{\theta}%
_{b},...,\boldsymbol{\theta}_{B})^{\prime}$. The symbol $p_{b}^{\ast}$
represents (the ordinate of) a kernel of the corresponding conditional
$p_{b}(\cdot|\cdot).$

\begin{algorithm}
	\caption{MH-within-Gibbs Algorithm}
	\label{MHwithinGibbsalg1}
	\begin{algorithmic}
		\STATE Specify an initial value $\bt^{(0)}$, a partition of the parameter set into $B$ mutually exclusive blocks, and a proposal distribution $q_b(\bt_b\mathbf{|y}_{1:T})$ for $b\in\{1,\dots,B\}$.
		\FOR{$i=1,\dots,M$}	
		\FOR{$b=1,\dots,B$}
		\STATE Draw $\bt_b^{c}\sim q_b(\bt_b\mathbf{|y}_{1:T})$
	    \STATE Compute the Metropolis-Hastings ratio:
		$$
		r=\frac{p_b^*(\bt_b^c|\bt_1^{(i)},\dots,\bt_{b-1}^{(i)},\bt_{b+1}^{(i-1)},\dots,\bt_{B}^{(i-1)},\mathbf{y}_{1:T})\times q_b(\bt_b^{(i-1)}\mathbf{|y}_{1:T})}{p_b^*(\bt_b^{(i-1)}|\bt_1^{(i)},\dots,\bt_{b-1}^{(i)},\bt_{b+1}^{(i-1)},\dots,\bt_{B}^{(i-1)},\mathbf{y}_{1:T})\times q_b(\bt_b^c\mathbf{|y}_{1:T})}
		$$
		\IF{$\mathcal{U}(0,1)<r$}
		\STATE Set $\bt^{(i)}_b=\bt^c$
		\ELSE
		\STATE Set $\bt_b^{(i)}=\bt_b^{(i-1)}$			
		\ENDIF
		\ENDFOR
		\ENDFOR
		\STATE Return a sample of draws from $p(\bt\mathbf{|y}_{1:T})$.
	\end{algorithmic}	
\end{algorithm}

The $b^{th}$ candidate density $q_{b}(\boldsymbol{\theta}_{b}\mathbf{|y}%
_{1:T})$ may be chosen to deliberately target the form of the $b^{th}$
conditional density, $p_{b}(\boldsymbol{\theta}_{b}\mathbf{|}%
\boldsymbol{\theta}_{1}^{(i)}\mathbf{,...,}\boldsymbol{\theta}_{b-1}%
^{(i)},\boldsymbol{\theta}_{b+1}^{(i-1)}\mathbf{,...,}\boldsymbol{\theta}%
_{B}^{(i-1)},\mathbf{y}_{1:T})$, in which case the algorithm may be referred
to as a `tailored' algorithm; otherwise $q_{b}(\boldsymbol{\theta}%
_{b}\mathbf{|y}_{1:T})$ may be chosen in a more automated fashion, such as in
a random-walk MH algorithm. The references cited in the text provide all details.

\subsection{ MH-within-Gibbs sampling in state space models\label{A3}}

The application of MH-within-Gibbs sampling within a state space setting is
qualitatively the same as described in Algorithm \ref{MHwithinGibbsalg1},
except that the joint set of unknowns is augmented to $(\boldsymbol{\theta}$,
$\mathbf{z}_{1:T})$, and decisions about partitioning need to be made for both
$\boldsymbol{\theta|}\mathbf{z}_{1:T}$ and $\mathbf{z}_{1:T}%
|\boldsymbol{\theta}$. Decisions about the blocking of $\mathbf{z}_{1:T}$ are
particularly important, given both the dimension of $\mathbf{z}_{1:T}$ and the
time-series dependence in the state process, as are matters of parameterizing
the state space model. We refer the reader to: \cite{shepard97} and
\cite{STRICKLAND2006} for illustrations of state blocking in which the block
sizes are selected randomly; and to \cite{fruhwirth:2004efficient} and
\cite{STRICKLAND2008} for explorations of the impact of parameterization on
the performance of the sampler.

\subsection{PMMH in state space models\label{A4}}

Early Bayesian treatments of non-linear state space models often exploited a
linear Gaussian approximation at some point, for the purpose of defining
candidate densities for (blocks of) $\mathbf{z}_{1:T}\,$(e.g.,
\citealp{kim1998svl}; \citealp{stroud2003}; \citealp{STRICKLAND2006}), thereby
enabling a Kalman filter-based `forward filtering, backward sampling'
algorithm (\citealp{carter:kohn:1994}; \citealp{fruhwirth-schnatter:1994}) to
be used to produce a candidate draw of (any particular block of)
$\mathbf{z}_{1:T}$, conditional on $\boldsymbol{\theta}$. As noted in Section
\ref{exact} (and in the review, \citealp{giordani2011bayesian}), more recent
approaches to such models have exploited PMMH principles instead. Algorithm
\ref{PMMHalg1} reproduces the algorithm in
\cite{andrieu:doucet:holenstein:2010} (Section 2.4.2 therein), adapted
slightly to match the notation of the current paper. To simplify the
exposition, the algorithm is presented for sampling the full vector
$\boldsymbol{\theta}$. In practice the algorithm would be modified to cater
for any blocking of $\boldsymbol{\theta}$.

\begin{algorithm}
	\caption{PMMH Algorithm}
	\label{PMMHalg1}
	\begin{algorithmic}
		\STATE Step 1: Initialization, $i=0$
		\begin{itemize}
			\item[(a)] Set $\bt^{(0)}$ arbitrarily and
			\item[(b)] Run an SMC algorithm targeting $p(\mathbf{z}_{1:T}|\mathbf{y}_{1:T},\bt^{(0)})$, sample $\mathbf{z}_{1:T}^{(0)}\sim \hat{p}(\mathbf{z}_{1:T}|\mathbf{y}_{1:T},\bt^{(0)})$ and let $\hat{p}(\mathbf{y}_{1:T}|\bt^{(0)})$ denote the marginal likelihood estimate.
		\end{itemize}
\STATE Step 2:
		\FOR{$i=1,\dots,M$}	
		\STATE (a) Draw $\bt^c\sim q(\bt|\mathbf{y}_{1:T},\bt^{(i-1)})$,\\
		\ \\
		\STATE (b) Run an SMC algorithm targeting $p(\mathbf{z}_{1:T}|\mathbf{y}_{1:T},\bt^{c})$, sample $\mathbf{z}_{1:T}^{c}\sim \hat{p}(\mathbf{z}_{1:T}|\mathbf{y}_{1:T},\bt^{c})$ and let $\hat{p}(\mathbf{y}_{1:T}|\bt^{c})$ denote the marginal likelihood estimate.\\
		\ \\
		\STATE (c) Compute the Metropolis-Hastings ratio:
		$$
		r=\frac{\hat{p}(\mathbf{y}_{1:T}|\bt^{c})p(\bt^{c})\times q(\bt^{(i-1)}|\mathbf{y}_{1:T},\bt^c)}{\hat{p}(\mathbf{y}_{1:T}|\bt^{{(i-1)}})p(\bt^{(i-1)})\times q(\bt^c|\mathbf{y}_{1:T},\bt^{(i-1)})}
		$$
		\IF{$\mathcal{U}(0,1)<r$}
		\STATE Set $\bt^{(i)}=\bt^c$, $\mathbf{z}_{1:T}^{(i)} = \mathbf{z}_{1:T}^{c}$, $\hat{p}(\mathbf{y}_{1:T}|\bt^{(i)}) = \hat{p}(\mathbf{y}_{1:T}|\bt^{c})$
		\ELSE
		\STATE Set $\bt^{(i)}=\bt^{(i-1)}$, $\mathbf{z}_{1:T}^{(i)} = \mathbf{z}_{1:T}^{(i-1)}$, $\hat{p}(\mathbf{y}_{1:T}|\bt^{(i)}) = \hat{p}(\mathbf{y}_{1:T}|\bt^{(i-1)})$
		\ENDIF
		\ENDFOR
		\STATE Return a sample of draws from $p(\bt|\mathbf{y})$.
	\end{algorithmic}	
\end{algorithm}

\subsection{ABC based on summary statistics\label{A5}}

The simplest (accept/reject) form of the ABC algorithm, as based on a chosen
vector of summaries, $\eta(\mathbf{y}_{1:T})$, proceeds via the steps in
Algorithm \ref{ABCalg2}, with the accepted draws of $\boldsymbol{\theta}$ used
to produce an estimate of $p_{\varepsilon}(\boldsymbol{\theta}\mathbf{|}%
\eta(\mathbf{y}_{1:T}))$, via kernel density methods. This posterior is
equivalent to $p(\boldsymbol{\theta}|\mathbf{y}_{1:T})$ if and only if
$\eta(\mathbf{y}_{1:T})$ is sufficient for conducting inference on
$\boldsymbol{\theta}$, and for $\varepsilon\rightarrow0$. Clearly, the very
problems for which ABC is required imply that sufficient statistics are not
available, and the requirement that $\varepsilon\rightarrow0$ is infeasible in
practice; so inference via ABC is only ever \textit{intrinsically}
approximate.\footnote{The notation $\mathbf{z}_{1:T}$ used in this section and
in Section \ref{A6} below is not to be confused with the use of $\mathbf{z}%
_{1:T}$ to denote a vector of latent variables elsewhere in the paper.}

\begin{algorithm}
	\caption{Accept/Reject ABC Algorithm Based on Summary Statistics}
	\label{ABCalg2}
	\begin{algorithmic}
	\FOR{$i=1,\dots,M$}	
	\STATE Simulate $\boldsymbol{\theta }^{(i)}$, $i=1,2,...,M$, from $p(%
	\boldsymbol{\theta })$, and artificial data $\mathbf{z}_{1:T}^{(i)}$ from $p(%
	\boldsymbol{\cdot }|\boldsymbol{\theta }^{(i)})$;
	\STATE Accept $\boldsymbol{\theta }^{(i)}$ if $d\{\eta (\mathbf{z}_{1:T}^{(i)}),\eta (%
\mathbf{y}_{1:T})\}\leq \varepsilon $, where $d\{\cdot ,\cdot \}$ denotes a
generic metric and $\varepsilon >0$ a pre-specified
tolerance parameter.
	\ENDFOR
	\end{algorithmic}	
\end{algorithm}

\subsection{BSL based on summary statistics\label{A6}}

BSL mimics ABC in targeting a posterior for $\boldsymbol{\theta}$ that
conditions on a vector of summaries $\eta(\mathbf{y}_{1:T})$, rather than the
full data set $\mathbf{y}_{1:T}$; however the summaries play a different role
in the algorithm. Once again with reference to the simplest version of the
algorithm, the steps of the BSL-MCMC algorithm are as given in Algorithm
\ref{BSLalg1}. Note that, for a given $\boldsymbol{\theta}$, the draws
$\mathbf{z}_{(j)1:T}\sim i.i.d.$ $p(\boldsymbol{\cdot}|\boldsymbol{\theta})$,
$j=1,\dots,m$, are used to estimate $\mu(\boldsymbol{\theta})$ and
$\Sigma(\boldsymbol{\theta})$ as $\mu_{m}(\boldsymbol{\theta})=\frac{1}{m}%
\sum_{j=1}^{m}\eta(\mathbf{z}_{(j)1:T})$ and $\Sigma_{m}(\boldsymbol{\theta
})=\frac{1}{m-1}\sum_{j=1}^{m}\mathcal{(}\eta(\mathbf{z}_{(j)1:T})-\mu
_{m}(\boldsymbol{\theta}))(\eta(\mathbf{z}_{(j)1:T})-\mu_{m}%
(\boldsymbol{\theta}))^{\prime}$. The $M$ draws of $\boldsymbol{\theta}$ are
used to produce an estimate of $p(\boldsymbol{\theta}\mathbf{|}\eta
(\mathbf{y}_{1:T}))$ via kernel density methods.

\begin{algorithm}
	\caption{BSL-MCMC Algorithm}
	\label{BSLalg1}
	\begin{algorithmic}
		\FOR{$i=1,\dots,M$}	
		\STATE Draw $\bt^\ast\sim q(\bt|\bt^{(i-1)})$
		\STATE Produce $\mu_m(\bt)$ and $\Sigma_m(\bt)$ using $j=1,\dots,m$ independent model simulations at $\boldsymbol{\theta }^\ast$
		\STATE Compute the synthetic likelihood $L^\ast=\mathcal{N}\left[ \eta (\mathbf{y});\mu _{m}(\boldsymbol{\theta }^\ast),\Sigma
		_{m}(\boldsymbol{\theta }^\ast)\right]$ and $L^{(i-1)}$ defined in a corresponding manner
		\STATE Compute the Metropolis-Hastings ratio:
		$$
		r=\frac{L^\ast\pi(\bt^\ast)q(\bt^{(i-1)}|\bt^\ast)}{L^{(i-1)}\pi(\bt^{(i-1)})q(\bt^\ast|\bt^{(i-1)})}
		$$
		\IF{$\mathcal{U}(0,1)<r$}
		\STATE Set $\bt^{(i)}=\bt^\ast$, $\mu_m(\bt^{(i)})=\mu_m(\bt^\ast)$ and $\Sigma_m(\bt^{(i)})=\Sigma_m(\bt^\ast)$	
		\ELSE
		\STATE Set $\bt^i=\bt^{(i-1)}$, $\mu_m(\bt^{(i)})=\mu_m(\bt^{(i-1)})$ and $\Sigma_m(\bt^{(i)})=\Sigma_m(\bt^{(i-1)})$			
		\ENDIF
		\ENDFOR
	\end{algorithmic}	
\end{algorithm}

\subsection{VB\label{A7}}

VB seeks the best approximation to $p(\boldsymbol{\theta}|\mathbf{y}_{1:T})$
over a `variational family' of densities $\mathcal{Q}$, with generic element
$q(\boldsymbol{\theta})$. Typically this proceeds by minimizing the
Kullback-Leibler (KL) divergence between $q(\boldsymbol{\theta})$ and
$p(\boldsymbol{\theta}|\mathbf{y}_{1:T})$, which produces the variational
approximation as
\begin{equation}
q^{\ast}(\boldsymbol{\theta}):=\argmin_{q(\boldsymbol{\theta})\in\mathcal{Q}%
}\text{KL}\left[  q(\boldsymbol{\theta})|p(\boldsymbol{\theta}|\mathbf{y}%
_{1:T})\right]  , \label{opt_1}%
\end{equation}
where
\begin{equation}
\text{KL}\left[  q(\boldsymbol{\theta})|p(\boldsymbol{\theta}|\mathbf{y}%
_{1:T})\right]  =\mathbb{E}_{q}[\log(q(\boldsymbol{\theta}))]-\mathbb{E}%
_{q}[\log(p(\boldsymbol{\theta},\mathbf{y}_{1:T}))]+\log(p(\mathbf{y}_{1:T}))
\label{kl}%
\end{equation}
{and }$p(\boldsymbol{\theta},\mathbf{y}_{1:T})=p(\mathbf{y}_{1:T}%
|\boldsymbol{\theta})p(\boldsymbol{\theta})$. Given that {the unknown
normalizing constant }$\log(p(\mathbf{y}_{1:T}))${ in (\ref{kl}) does not
depend on $q$, the (infeasible) optimization problem in \eqref{opt_1}, is
replaced by the equivalent (and feasible) optimization problem:
\begin{equation}
q^{\ast}(\boldsymbol{\theta}):=\argmax_{q(\boldsymbol{\theta})\in\mathcal{Q}%
}\left\{  \mathbb{E}_{q}[\log(p(\boldsymbol{\theta},\mathbf{y}_{1:T}%
))]-\mathbb{E}_{q}[\log(q(\boldsymbol{\theta}))]\right\}  , \label{vb_1}%
\end{equation}
with the so-called evidence lower bound (ELBO) defined as:}%
\begin{equation}
\text{ELBO}[q(\boldsymbol{\theta})]:=\mathbb{E}_{q}[\log(p(\boldsymbol{\theta
},\mathbf{y}_{1:T}))]-\mathbb{E}_{q}[\log(q(\boldsymbol{\theta}))].
\label{elbo}%
\end{equation}

The usefulness of VB is that, for certain models, $p(\mathbf{y}_{1:T}%
|\boldsymbol{\theta})$, and certain choices of $\mathcal{Q}$, the optimization
problem in (\ref{vb_1}) {can be solved efficiently using} various numerical
algorithms. Most notably, for problems in which the dimension of the unknowns,
{and possibly that of }$\mathbf{y}_{1:T}$ {also}, {is large, the production of
}$q^{\ast}(\boldsymbol{\theta})$ {is \textit{much} faster (often orders of
magnitude so) than producing an estimate of }$p(\boldsymbol{\theta}%
|\mathbf{y}_{1:T})$ (and any associated quantities, including predictives)
{via simulation} (\citealp{braun2010variational}; \citealp{kabisa2016online};
\citealp{wand2017fast}; \citealp{koop2018variational}). {The relationship
between (\ref{kl}) and (\ref{elbo}), plus the fact that }${KL[}\cdot]\geq0$,
{also means }ELBO$[q^{\ast}(\boldsymbol{\theta})]${ is a lower bound on the
logarithm of the `evidence', or marginal likelihood, }$p(\mathbf{y}_{1:T})$;
hence the abbreviation `ELBO'.{ }

{Different VB methods are defined by both the choice of }$\mathcal{Q}$ and the
manner in which the optimization is implemented, and we refer the reader to
\cite{ormerod2010explaining}, \cite{blei2017variational}, and
\cite{zhang2018advances}, for reviews, including algorithmic details for
specific VB methods.

\subsection{INLA\label{A8}}

\cite{rue:martino:chopin:2009} adapted the very early approximation method of
\cite{laplace:1774} to approximate posteriors (and associated quantities) in
the latent Gaussian model class, which encompasses a large range of --
potentially high-dimensional -- models, including the non-Gaussian state space
models that feature heavily in economics and finance. In brief,
\citeauthor{rue:martino:chopin:2009} use a series of \textit{nested }Laplace
approximations, allied with low-dimensional numerical \textit{integration},
denoting their method by \textit{integrated nested Laplace approximation
}(INLA) as a result. As with VB, INLA eschews simulation for optimization,
exploiting bespoke numerical algorithms designed for the specific (albeit
broad) model class. We refer the reader to \cite{rue:martino:chopin:2009},
\cite{Rue2017}, \cite{martino2019integrated}, and \cite{wood2019simplified}
for implementation details.
\end{document}